\documentclass[iop]{emulateapj}
\usepackage{graphicx}
\usepackage{epstopdf}
\usepackage{natbib}
\usepackage{xspace}
\usepackage{color}

\newcommand{\sod}	{SO$_2$}

\newcommand{\meta}	{CH$_3$OH}

\newcommand{\htcop}	{H$^{13}$CO$^+$}
\newcommand{\hcqn}	{HC$^{15}$N}
\newcommand{\DE}	{	CH$_3$OCH$_3$}
\newcommand{\dr} 	{DR~21(OH)}

\newcommand{\kms}	{~km~s$^{-1}$\xspace}

\newcommand{\msun}	{~$M_{\sun}$\xspace}
\newcommand{\lsun}	{~$L_{\sun}$\xspace}
\newcommand{\cmd}	{~cm$^{-2}$\xspace}
\newcommand{\cmt}	{~cm$^{-3}$\xspace}
\newcommand{\vlsr}	{~$v_{\rm LSR}$\xspace}
\newcommand{\ie}	{i.\,e.,}

\newcommand{\hii}	{\ion{H}{2}}

\shorttitle{Magnetic Fields in DR 21(OH)} 
\shortauthors{Girart et al.}

\begin{document}

\title{DR~21(OH): a highly fragmented, magnetized, turbulent dense core}

\author{J.M. Girart\altaffilmark{1}}
\author{P. Frau\altaffilmark{1,2}}
\author{Q. Zhang\altaffilmark{3}}
\author{P. M. Koch\altaffilmark{4}}
\author{K. Qiu\altaffilmark{5}}
\author{Y.-W. Tang\altaffilmark{4}}
\author{S.-P. Lai\altaffilmark{4,6}}
\author{P.T.P. Ho\altaffilmark{4}}

\altaffiltext{1}{Institut de Ci\`encies de l'Espai, (CSIC-IEEC), Campus UAB,
Facultat de Ci\`encies, C5p 2, 08193 Bellaterra, Catalonia;
girart@ice.cat}
\altaffiltext{2}{Observatorio Astron\'omico Nacional, Alfonso XII, 3
E-28014 Madrid, Spain}
\altaffiltext{3}{Harvard-Smithsonian Center for Astrophysics, 
60 Garden Street, Cambridge, MA 02138, USA}
\altaffiltext{4}{Academia Sinica Institute of Astronomy and Astrophysics, 
P.O. Box 23-141, Taipei 10617, Taiwan}
\altaffiltext{5}{School of Astronomy and Space Science, Nanjing University, 
Nanjing 210093, China} 
\altaffiltext{6}{Institute of Astronomy and Department of Physics, National 
Tsing Hua University, 101 Section 2 Kuang Fu Road, Hsinchu 30013, Taiwan}

\begin{abstract}
We present high angular resolution observations of the massive star forming 
core DR21(OH) at 880~$\mu$m using the Submillimeter Array (SMA). 
The  dense core exhibits an overall velocity gradient in a Keplerian-like
pattern,  which breaks at the center of the core where  SMA 6 and SMA 7 are
located.  The dust polarization shows a complex magnetic field, compatible with
a  toroidal configuration.  This is in contrast with the large, parsec--scale 
filament that surrounds the core, where there is a smooth magnetic field.   The
total magnetic field strengths in the filament and in the core are 0.9 and 
2.1~mG, respectively. We found evidence of magnetic field diffusion at the core 
scales, far beyond the expected value for ambipolar diffusion. It is possible 
that the diffusion arises from fast magnetic reconnection in the presence of 
turbulence.  The dynamics of the \dr\ core appear to be controlled
energetically  in equal parts by the magnetic field, magneto--hydrodynamic (MHD)
turbulence  and the angular momentum.  The effect of the angular momentum (this
is a fast  rotating core) is probably causing the observed toroidal field
configuration.  Yet, gravitation overwhelms all the forces, making this a clear
supercritical  core with a mass--to--flux ratio of $\simeq 6$ times the critical
value.  However, simulations show that this is not enough for the high level of 
fragmentation observed at 1000~AU scales.  Thus, rotation and outflow feedback 
is probably the main  cause of the observed fragmentation.
\end{abstract}

\keywords{ISM: individual objects (\dr) -- ISM: magnetic fields -- polarization --
stars: formation -- submillimeter: ISM -- techniques: polarimetric}

\section{Introduction}\label{intro}

DR 21(OH), also known as W75, is a well-studied high-mass star forming region, 
located inside the Cygnus X molecular cloud complex \citep{Downes66, Motte07,
Reipurth08}. It is located in a dense, 4~pc long, DR~21 filamentary  ridge,
active in star formation, with global infall motions \citep{Harvey86, Vallee06,
Csengeri11, Schneider10, Hennemann12}. The distance to the DR21 region has been 
recently re-estimated through trigonometric parallaxes of  masers, 
$1.50\pm0.08$~kpc \citep{Rygl12}, a factor two lower than the previous 
estimations. We have re-estimated some physical parameters  given in previous 
works taking into account the new distance.  

High angular resolution continuum observations show that the \dr\ core is 
formed by two bright sources, MM~1 and MM~2 \citep{Woody89, Lai03}.  But 
recent subarcsecond 
angular resolution observations show that these two sources split into a cluster
of  dusty sources at scales of 1000~AU \citep{Zapata12}. Chemical analysis of
the two  main clumps show that MM~1 is more evolved than  MM~2
\citep{Mookerjea12},  which is in agreement with the mid-IR images that show
bright emission from MM~1  but no emission toward MM~2
\citep[e.g.,][]{Araya09}.  The total bolometric luminosity  of \dr\ is 
$1.6Ê\times 10^4$~\lsun\ \citep{Jakob07}. \dr\ shows very active and powerful 
dense outflows, traced not only by  CO but also by SiO, CH$_3$OH, H$_2$CO and 
H$_2$CS  \citep{Lai03,Minh11,Zapata12}.   It also shows a rich variety of masers
from  molecules such as OH, CH$_3$OH (class I and II), water and HCO$^+$ 
\citep[e.g.,][]{Matthews86, Batrla88, Plambeck90,  Harvey08, Araya09, Fish11, 
Hakobian12}.  

Magnetic fields at large parsec scales have been mapped through single-dish
polarimetric observations \citep{Minchin94,Glenn99,Vallee06,Kirby09} revealing
a  relatively uniform magnetic field orientation.  Higher angular resolution
interferometric  observations at millimeter wavelengths \citep{Lai03} resolve the
magnetic field in the core.  Zeeman observations of the CN line reveals a
magnetic field strength in the line--of--sight of $B_{\rm los} \simeq
0.4$--0.7~mG \citep{Crutcher99}.

Polarization observations with the Submillimeter Array (SMA) have been 
successfully carried out since
2006.  In the earlier evolutionary stage of molecular clouds (e.g. collapsing
phase),  hourglass--like magnetic field lines have been detected in both
low--mass star--forming regions \citep[NGC~1333 IRAS~4A,
IRAS16293$-$2422:][]{Girart06,  RaoEtAl2009}, and high-mass star-forming regions
\citep[G31.41+0.31, W51~e2  and W51~North:][]{Girart09,  Tang09b, Tang13},
suggesting magnetic--field--regulated  gravitational collapses.  In contrast,
the influences of stellar feedbacks on the magnetic  field are seen in more
evolved ultra compact HII regions \citep[G5.89$-$0.39,  
NGC~7538~IRS1:][]{Tang09a, frau13} and in the Orion BN/KL region 
\citep{Tang10}.  Very recently, CARMA has also started to carry out
polarimetric  observations of dust emission \citep{Hull13}.

In this paper we present SMA spectro-polarimetric observations carried out at
345~GHz toward DR~21(OH). Here, we focus on the dust polarization observations. 
Additional data of selected molecular lines are included to better understand
the  overall properties of this region.  Section~\ref{obs} briefly describes the
observations and  Section~\ref{res}  presents the results of the observations. A
statistical analysis of the  dust polarization is presented in Section~\ref{ana}.
Section~\ref{dis} contains the discussion.  Finally, in Section~6 we draw the 
main conclusions.

\section{Observations}\label{obs}

The observations were taken with the SMA \citep{Ho04} between 2011 June and October
in different array configurations.  Table~\ref{T0} lists the observation 
dates, and for each  date the configuration used, the number of antennas, the
total  amount of time on-source and the polarization calibrators. For all
observations but  the one from June 30, a single receiver was used around 345
GHz, with a total  bandwidth  of 4~GHz per sideband.  The receiver was tuned to
cover the  332.1--336.0 and 344.1--348.0~GHz frequencies in the lower (LSB) and
upper  sideband (USB), respectively.  For the observation on June 30 (in
subcompact  configuration), the dual--receiver mode was used, tuning the 345 and
400 GHz  receivers to the same frequency, which covered the 334.0--335.9 and 
344.0--345.9~GHz frequencies in the LSB and USB,  respectively. The phase 
center was 
$\alpha$(J2000.0)$=20^{\rm h}39^{\rm m}01\fs20$ and  
$\delta$(J2000.0)$= $42$\degr22\arcmin 48\farcs50$.  
The correlator provided a spectral resolution of about 0.8~MHz (i.e.,
0.7~km~s$^{-1}$ at 345~GHz) for the single-receiver mode.  The gain calibrator 
was MWC349A.  The bandpass calibrator was the same as the polarization 
calibrator  (see Table~\ref{T0}).  The absolute flux scale was determined from 
observations of Ceres and Callisto.  The flux uncertainty was estimated to be 
$\sim20$\%.  The data were reduced using the IDL MIR and MIRIAD software 
packages.

%%% OBSERVATIONAL PARAMETERS. I
\begin{deluxetable}{lclcl}
\tablecaption{Observational Parameters\label{T0}}
\tablewidth{0pt}
\tablehead{
\multicolumn{2}{c}{} & 
\multicolumn{1}{c}{$\!\!\!\!\!\!\!\!\!\!\!\!$Number} &
\multicolumn{1}{c}{$\!\!\!\!$On-source} &
\multicolumn{1}{c}{} 
\\
\multicolumn{1}{c}{Date of} & 
\multicolumn{1}{c}{} & 
\multicolumn{1}{l}{$\!\!$of} &
\multicolumn{1}{c}{$\!\!\!\!$Observing}  & 
\multicolumn{1}{l}{Polariz.} 
 \\
\multicolumn{1}{c}{Observations} & 
\multicolumn{1}{c}{$\!\!\!\!$Configuration} & 
\multicolumn{1}{c}{$\!\!\!\!\!\!\!$Antennas} &
\multicolumn{1}{c}{$\!\!\!\!$Time} & 
\multicolumn{1}{l}{Calibrator} 
}
\startdata
2011 Jun 30 & Subcompact	& 7 & 0.90 hr & 3C454.3 \\
2011 Jun 21 & Compact 	& 8 & 0.21 hr & 3C279\\
2011 Jul 13 & Compact 	& 7 & 0.41 hr & 3C279 \\
2011 Oct 17 & Compact 	& 7 & 0.78 hr & 3C84 \\
2011 Jul 18 & Extended	& 8 & 0.35 hr & 3C454.3 \\
2011 Jul 20 & Extended 	& 8 & 0.35 hr & 3C279 \\
2011 Jul 21 & Extended 	& 8 & 0.97 hr & 3C279 \\
2011 Jul 23 & Extended 	& 8 & 0.76 hr & 3C279 \\
2011 Sep 03 & Very extended & 8 & 0.26 hr & 3C84 
\enddata
\end{deluxetable}

The SMA conducts polarimetric observations by cross-correlating  circular
polarizations (CP). The CP is produced by inserting quarter wave  plates in
front of the receivers which are inherently linearly polarized. A  detailed
description of the instrumentation techniques as well as calibration  issues is
discussed in  \citet{Marrone08} and \citet{Marrone06}.  We found  polarization
leakages between 1\% and 2\% for the USB, while the LSB leakages  were between 2\%
and 4\%. These leakages were measured to an accuracy of 0.1\%
\citep{Marrone08}.

\begin{figure}[h]
\begin{center}
\includegraphics[width=\columnwidth]{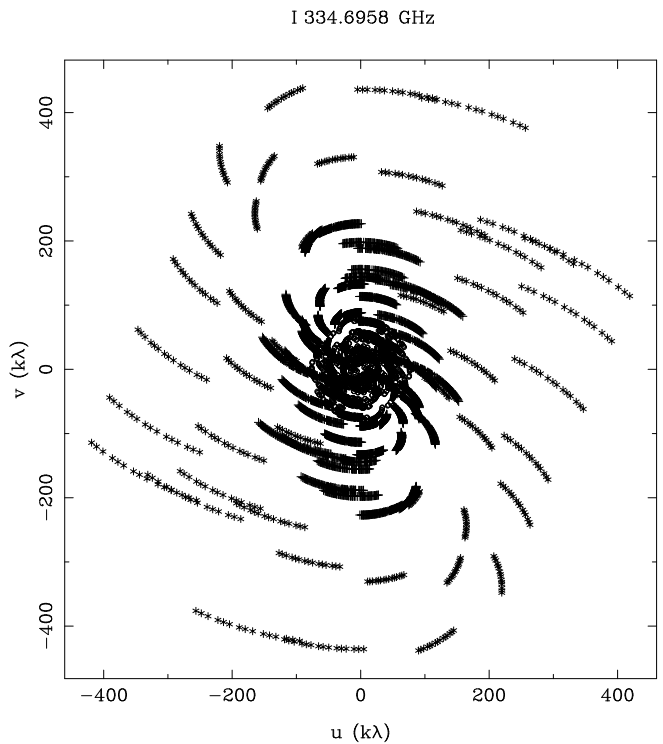}
\caption{Visibility coverage of the SMA observations, which includes all the
configurations (subcompact, compact, extended and very extended).
}
\label{Fuvcover}
\end{center}
\end{figure}

%%% OBSERVATIONAL PARAMETERS. II
\begin{deluxetable}{cccccl}
\tablecaption{Mapping Parameters\label{T1}}
\tablewidth{0pt}
\tablehead{
\multicolumn{1}{c}{} & 
\multicolumn{2}{c}{$u,v$} & 
\multicolumn{1}{c}{Synthesized} &
\multicolumn{2}{c}{rms Noise} 
\\
\cline{2-3}
\cline{5-6}
\multicolumn{1}{c}{Configu-} & 
\multicolumn{1}{c}{Range} & 
\multicolumn{1}{c}{Taper} &
\multicolumn{1}{c}{Beam} &
\multicolumn{1}{r}{Stokes $I$}  &
\multicolumn{1}{r}{Pol}  
\\
\multicolumn{1}{l}{ration\tablenotemark{a} } & 
\multicolumn{1}{c}{k$\lambda$} & 
\multicolumn{1}{c}{$''$\tablenotemark{b}} &
\multicolumn{1}{c}{{\it FWHM, PA}\tablenotemark{c}} &
\multicolumn{2}{c}{mJy~beam$^{-1}$} 
}
\startdata
SCE	& 0,90	& 3.0& $3\farcs86$$\times$$3\farcs42$, $56\arcdeg$& 20 & 3.1 \\
SCEV	& 0,450	& 0.4& $1\farcs51$$\times$$1\farcs21$, $82\arcdeg$&  9 & 1.2 \\
CEV	&30,450 & 0.0& $0\farcs87$$\times$$0\farcs65$, $89\arcdeg$&  3 & 1.2
\enddata
\tablenotetext{a}{S: subcompact, C: compact, E: extended, V: very extended}
\tablenotetext{b}{Gaussian taper applied to the visibility data in image
units. }
\tablenotetext{c}{$FWHM$: full-width at half-maximum, $PA$: position angle}
\end{deluxetable}

Self-calibration was performed using the Stokes $I$ continuum data for each 
antenna configuration independently.  The derived gain solutions were applied 
to the molecular line data.  The whole data set includes all the different SMA 
configurations, covering a wide range of visibilities (from 6 up to
450~k$\lambda$).  Each configuration is designed to have a relatively uniform
density of visibilities.  This implies that the combination of several
configurations results in a coverage of  visibilities with a heterogeneous
density (see  Figure~\ref{Fuvcover}). Therefore, to   take advantage of all the
information that the whole data contain, maps with different  visibility
weightings and $u,v$ coverages were used.  Table~\ref{T1} lists the basic 
parameters of the resulting different maps presented in this paper: $u,v$
coverage,  weighting, configuration, synthesized beam, spectral resolution and
resulting $rms$  noise.   The map at $3''$ angular resolution (SCE as defined
in Table~\ref{T1}) was made to compare our results with the previous BIMA 
polarimetric maps \citep{Lai03}. The map at $1''$ angular resolution 
(Table~\ref{T1}: SCEV) takes advantage of the full visibility coverage at the
highest angular resolution and the best sensitivity for the polarization. The
subarcsecond map  (Table~\ref{T1}: CEV) avoids the larger scale
dust emission to trace the magnetic field at scales of few thousands AU. This is
done by excluding the shortest baselines.  The significantly higher $rms$ noise 
of the SCEV and CEV Stokes $I$ maps is due to the limited dynamic range of
the SMA (the shortest visibilities have strong Stokes $I$ amplitudes).  
Table~\ref{T2} gives the transitions, frequency, lower energy level of the 
molecular lines presented in this paper, as well as the $rms$  noise for  channel 
maps with a velocity width of 1.5~\kms.  The figures were
created using the GREG package (from the  GILDAS\footnote{GILDAS data reduction
package is available at  http://www.iram.fr/IRAMFR/GILDAS}  software).

\begin{figure*}[]
\includegraphics[width=8.5cm,angle=-90]{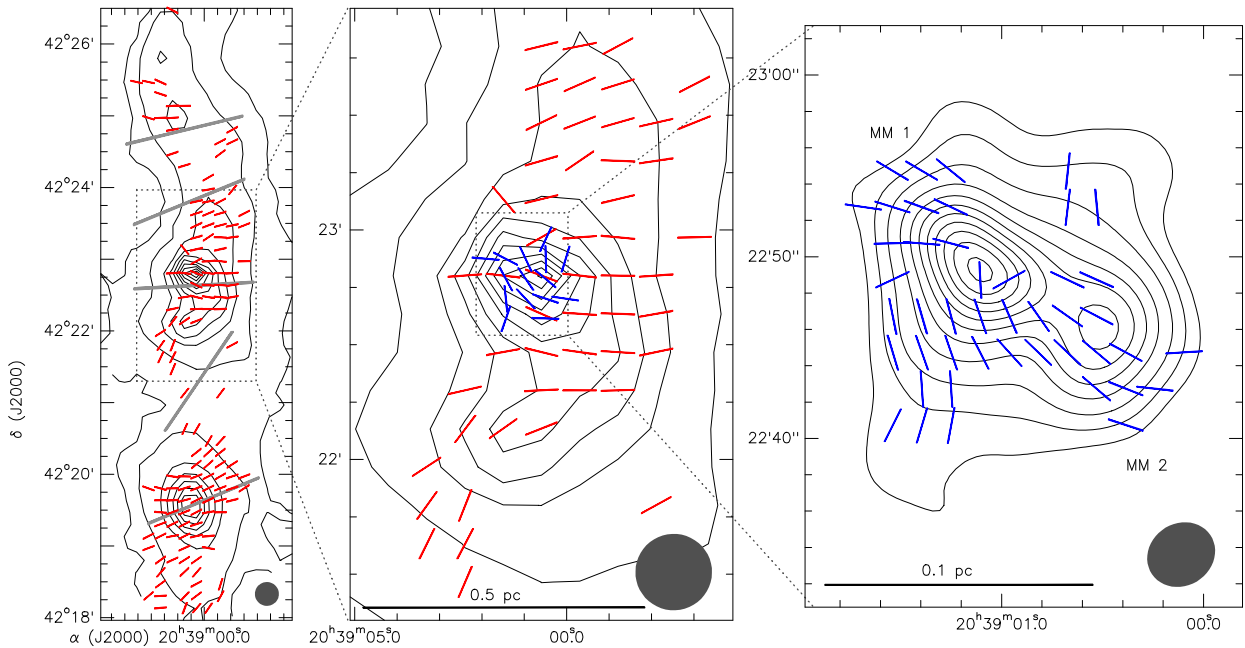}
\caption{
{\it Left panel:} contour map of the dust emission at 850~$\mu$m towards the
DR21  region, overlapped with the $B$ segments (red bars) obtained with the JCMT
SCUBA  polarimeter, SCUPOL  \citep{Vallee06, Matthews09}. \dr\ is the brightest
core  located at the center  of the panel. The angular resolution of the map is
$20''$.  Contours are 4\%, 8\%, 18\%, \nodata, 98\% of the peak.  
{\it Middle panel:} zoom-in of the previous panel toward \dr. The blue bars
show the SMA B segments obtained at an angular resolution of $10''$.
{\it Right panel:}  contour map of the dust emission at 880~$\mu$m obtained
with the SMA (SCE map as defined in Table~\ref{T1}).  This panel shows the 
same map as Fig.~\ref{Fbima}. Blue bars depict the B segments. 
}
\label{Fscuba}
\end{figure*}

\section{Results}\label{res}

In this section we describe the results obtained with the SMA. For the total and
polarized dust emission, we also present the single--dish James Clerk Maxwell 
telescope (JCMT) data obtained
with SCUPOL\footnote{These data were obtained  from  the SCUBA Polarimeter
Legacy Catalog compiled by \citet{Matthews09} and  were previously published by
\citet{Vallee06}.}.  
This allows us to study the magnetic fields from parsec to few thousandths of a 
parsec scale.  Hereafter we define three different physical structures observed
at different scales: The DR 21 filament, which is the parsec-long structure
where \dr\ is embedded \citep{Vallee06}; The \dr\ core, a 0.1~pc structure that
is resolved into two continuum peaks, MM~1 and MM~2  \citep{Lai03}  when
observed at an arcsecond angular resolution; The  substructures detected in the
millimeter/submillimeter dust continuum emission maps  at scales of 1000~AU 
\citep[sources SMA
1--9 by][]{Zapata12} will be  referred as condensations.

\subsection{Dust Emission and Magnetic Fields: from Parsec to Sub-parsec 
Scales\label{Sdust}}

Previous observations have shown that \dr\ is embedded in a 4 pc long dense  and
massive ($1.5 \times 10^4$~\msun) filament  extending in the north-south
direction  \citep{Vallee06, Hennemann12}.  The filament harbors other star
forming cores, such  as the well known \hii\ region DR 21 main  \citep{Vallee06,
Hennemann12}.  Figure~\ref{Fscuba} shows the submillimeter dust emission arising  from
the filament and the magnetic field that threads the filament. DR 21 main and
\dr\ are the bright  cores located at the south and at the center of the
filament,  respectively. The gray long bars shown in this figure represent the
average  direction of the  magnetic field in different sections of the
filament.  Interestingly,  the magnetic  field direction in the plane of the sky
is close to the East--West direction, and thus  almost perpendicular  to the
filament. The exception is a small region with weak  polarization between \dr\
and DR21 main, where the direction flips to a position  angle of
$\simeq$146\arcdeg. The variation of the direction along the filament  occurs
smoothly.  Around DR 21  main the magnetic field configuration is  compatible
with the hourglass  morphology \citep{Kirby09}. There are other reports of
massive filaments with  magnetic fields perpendicular to the filament 
\citep[e.g., G14.225,][]{busquet13}.

The zoom--in of the single--dish polarization map toward the filament around
\dr\  (the middle panel of Figure~\ref{Fscuba}, red bars) shows that the magnetic field is
relatively uniform  and mostly in the east-west direction. This pattern of the
magnetic field is  in agreement with CO J=2-1 and 1-0 polarimetric  data derived
with BIMA  \citep{Lai03, Cortes05}, which trace the low density molecular gas, 
$n({\rm H}_2)\simeq10^2$~\cmd. In contrast, the SMA polarization map at an 
angular resolution of $3''$ reveals field orientations much less uniform (see
the right  panel of Figure~\ref{Fscuba}). To properly compare the SMA and SCUBA 
polarization maps,  we convolved the $3''$ SMA map with a Gaussian to degrade 
the angular resolution up to $10''$.   The resulting map, shown in the central
panel  of Figure~\ref{Fscuba}, reveals that the magnetic field derived from the
SMA  at this angular resolution (blue segments in this figure) is still less
uniform, even  though some of the magnetic field segments are roughly aligned in
the E-W  direction, the direction of the filament component. It is 
important to remark that the SMA  filters out
the  large-scale component from the dust total and polarized intensity. 
Therefore, the  SMA is more sensitive to the small-scale magnetic field within
the  core, whereas the single-dish map is more sensitive to the total column
density of  dense molecular gas, thereby to the large-scale component.   We,
thus, do not  necessarily expect to recover the SCUBA field morphology after
convolving the  SMA data.

\subsection{Dust Emission and Magnetic Fields: from 20,000 to 1000~AU 
Scales\label{Sdust2}}

\begin{figure*}[h]
\begin{center}
\includegraphics[width=11.5cm]{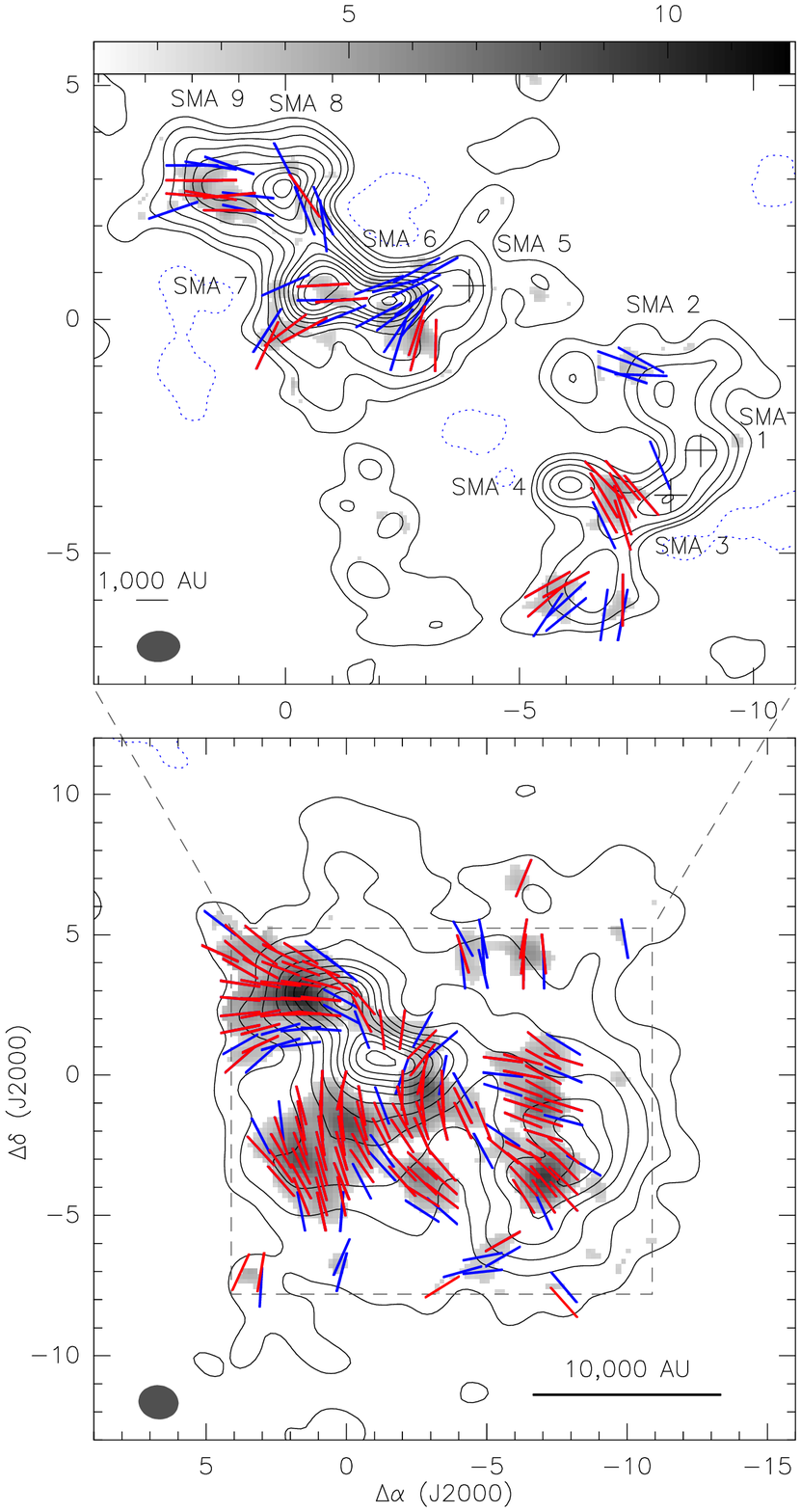}
\caption{Contour map of the dust emission at 880~$\mu$m, overlapped with the
gray scale  intensity of the dust linear polarized emission and the B segments.
{\it Bottom panel:} Images with an angular resolution of $\simeq 1\farcs3$ 
obtained using all the configurations (SCEV map as defined in Table~\ref{T1}). 
{\it Top panel:} Images with an angular resolution of $\simeq 0\farcs75$  obtained 
using all the configurations but the subcompact one (CEV map as defined in 
Table~\ref{T1}). The red and blue segments show the magnetic
field segments with a significance  level of $\ga 3$--$\sigma$ and between 2.5
and 3~$\sigma$, respectively. The synthesized beam is shown in the bottom left
corner of each panel.  Contours in the two panels are 5\%, 10\%, 17\%, 27\%, 
\nodata, 97\% of the peak, 0.85  and 0.29~Jy~beam$^{-1}$ for the $1\farcs3$ and
$0\farcs75$ angular  resolution maps, respectively. 
}
\label{Fpol}
\end{center}
\end{figure*}

Observations at an angular resolution of $3\farcs6$ ($\simeq 5500$~AU, see
Figure~\ref{Fbima}) show that the millimeter continuum emission is  dominated by the dust
emission and arises from two main components, MM~1 and  MM~2 \citep{Woody89,
Lai03}. However, the higher angular resolution map  reveals clearly how this
region fragments in a significant way. First, the  map obtained using all the
visibilities and with an angular resolution of  $\simeq 1\farcs3$ (configuration
SCEV from Table~\ref{T1}; see also the bottom panel of Figure~\ref{Fpol})  shows that
MM~1  has split into two bright components, whereas MM 2 appears  elongated with
an arc-like morphology.  The fragmentation is more evident at  sub--arcsecond
scales ($\simeq 1000$~AU, as shown in the top panel of  Figure~\ref{Fpol}) for a
map obtained excluding the  shortest baselines   ($r_{\rm u,v} <
30$~k$\lambda$),  and thereby filtering some  of the extended  component that
appears in the $\simeq 1\farcs3$  map.  At this angular resolution,  the 880~
$\mu$m map is in agreement with the 1.4~mm map obtained by  \citet{Zapata12},
although the better sensitivity allows us to detect more  emission.  MM~1 splits
into four  bright sources (from west to east, SMA 6,   SMA 7, SMA 8 and SMA 9,
according to Zapata et al. 2012 nomenclature).   SMA~5 is not well resolved at
this angular resolution. SMA 6 and SMA 7 are the  sources located closer to the
center of the whole dense  molecular core.  MM 2 splits in several
components: a compact source  SMA 4, an elongated structure that contains SMA 1,
SMA 2 and SMA 3, and  possibly two additional components not previously
reported: one $2''$ south of SMA~4 and the other $2''$ east of SMA~2.  In brief,
\dr\ splits probably in more  than 10 sources and this constitutes an extreme case
of a highly fragmented dense  molecular core, according to a recent study
carried out at similar spatial scales  over a sample of 18 intermediate and
massive dense cores  \citep{Palau13}.

The total flux measured at 880~$\mu$m is $18.8\pm0.1$~Jy. To estimate the  mass 
we adopt a gas-to-dust ratio of 100 and a dust opacity of 1.5~cm$^2$~g$^{-1}$, 
which is approximately the expected  value for dust grains with thin dust
mantles at  densities of $\sim 10^6$~ \cmt  \citep{Ossenkopf94}. Assuming a
temperature of 30~K  \citep{Mayer73,Vallee06}, we then estimate the total mass
traced by the dust to be  150~\msun.  This value is a factor of 2 lower than the
mass derived from single--dish   measurements of the dust emission
\citep[350~\msun:][]{Motte07}, which is likely due  to the filtering effect of
the SMA\footnote{We convolved the dust continuum map with a Gaussian to
obtain an angular resolution of $14''$, which is the value of the JCMT beam. The
intensity measured at the peak of the convolved map is about 40\% lower than the 
value measured with the JCMT \citep{Vallee06}}.  To derive the averaged volume 
and column  densities in the whole \dr\
core, we use the full width at half maximum (FWHM) of the  dust emission at an
angular resolution of $3''$,  FWHM$\simeq 10\farcs4$. This value  yields an
average column and volume density of $1.6\times10^{24}$~\cmd and  $1.0
\times10^7$~\cmt, respectively.

\begin{figure}[]
\includegraphics[width=\columnwidth]{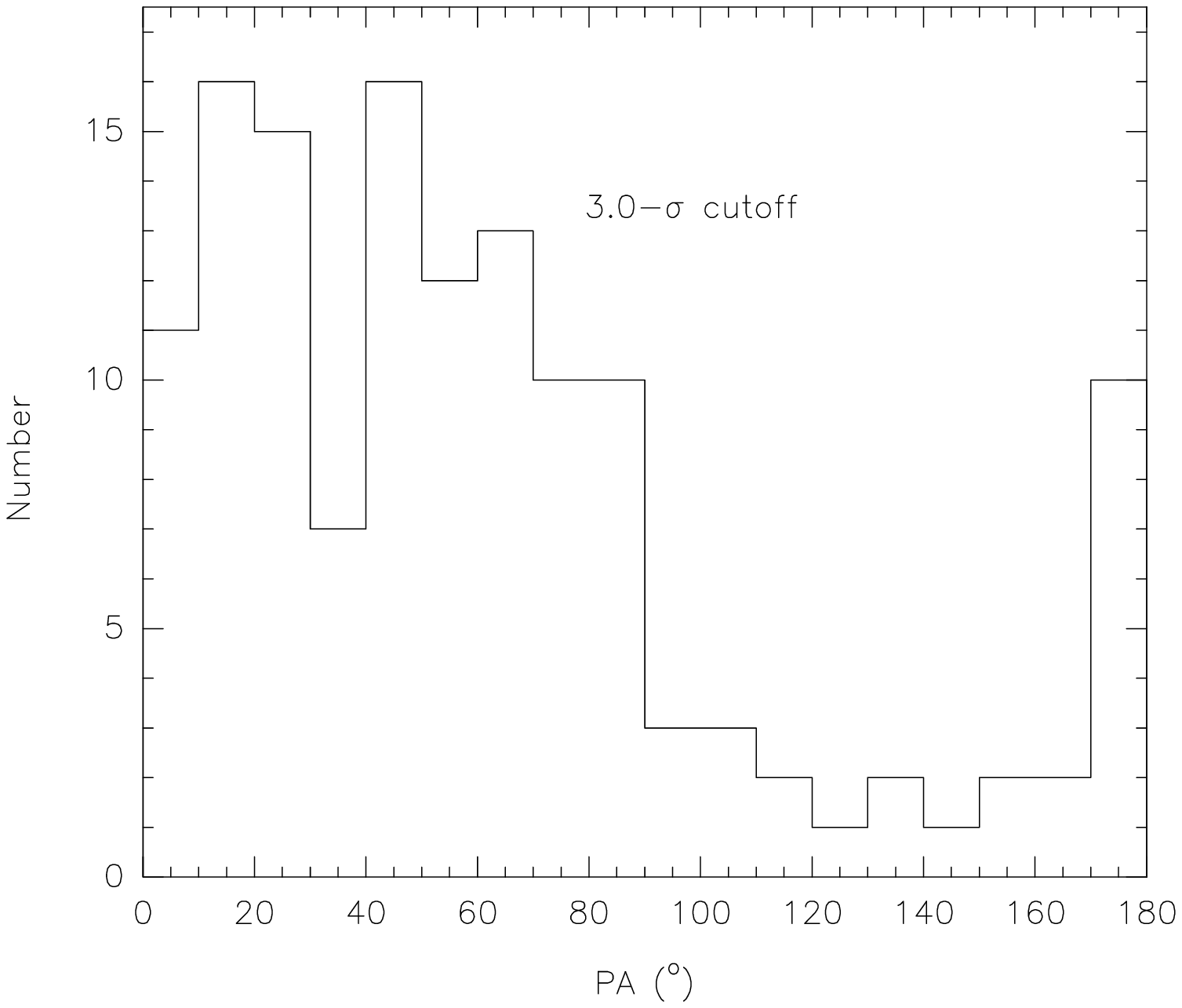}
\caption{Distribution of position angles of the magnetic field segments shown
in  the $1\farcs32$ angular-resolution map for a polarized emission cutoff of 
3.0--$\sigma$.}
\label{Fpol2}
\end{figure}

\begin{figure}[h]
\includegraphics[width=\columnwidth]{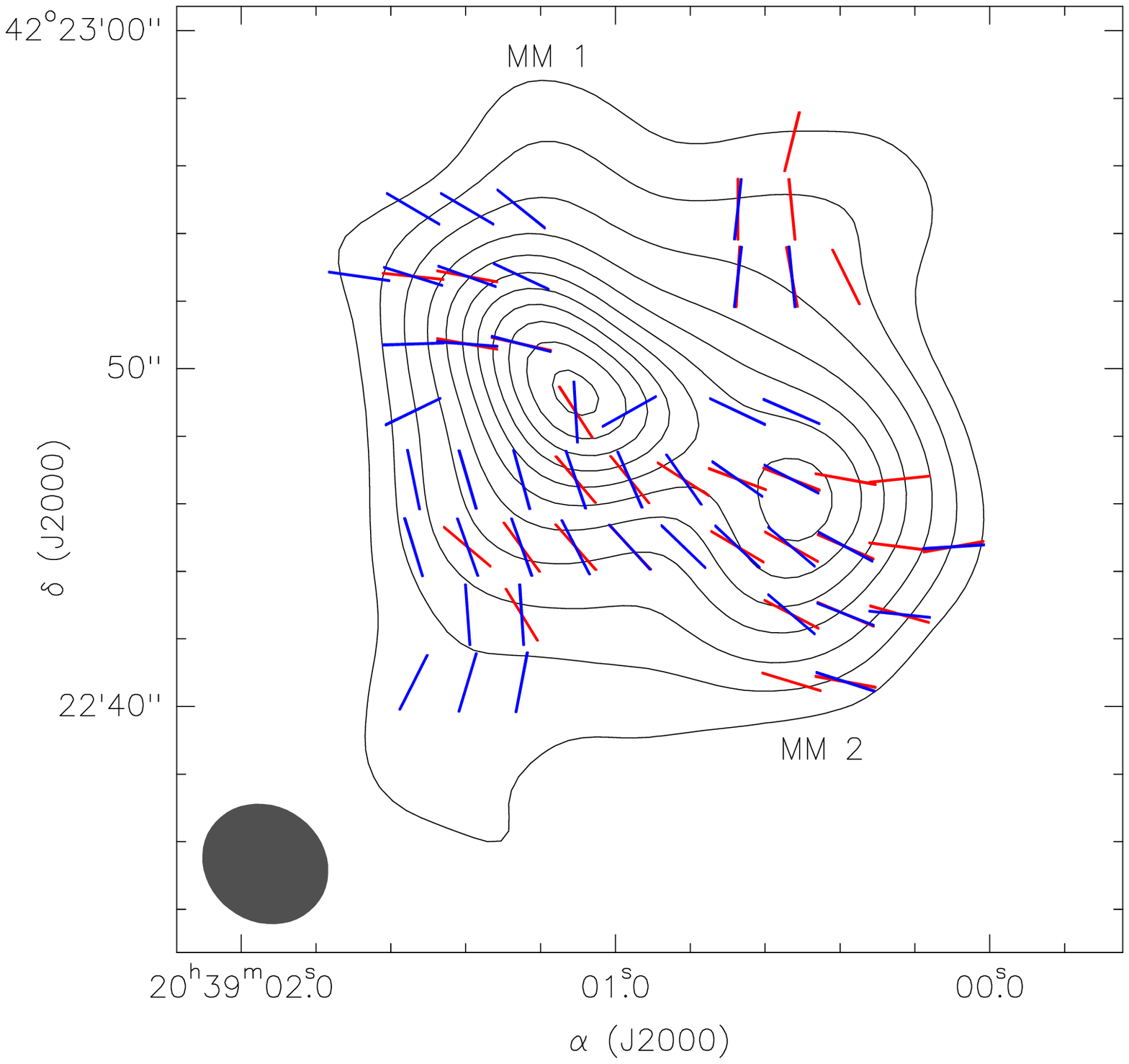}
\caption{Contour map of the SMA dust emission at 880~$\mu$m. The map  was
obtained from the combined data of the subcompact, compact  and extended
configurations with a Gaussian taper to match the BIMA  polarization maps by
\citet{Lai03}. The synthesized beam is shown in the  bottom left corner of the
panel. Overlapped with the contour maps are the B segments derived at a similar
angular resolution by BIMA (red segments; from Lai et al. 2003) and the SMA
(blue segments). 
}
\label{Fbima}
\end{figure}

Figure~\ref{Fpol2} shows the distribution of the angles of the magnetic field 
segments measured in the  $1\farcs3$ angular resolution  map
with a Nyquist  sampling and with a cutoff in the polarized emission of 
3--$\sigma$. The distribution shows a broad dispersion in the  0 to 80\arcdeg\
range without a clear main direction.  However, a visual inspection  of the
resulting magnetic field (see bottom panel of Figure~\ref{Fpol})  seems to show 
that there are two main directions of the magnetic field: (1) NE--SW  around MM
2  and east of MM 1 and (2) N--S in the northern part of  the dense core and South
of  MM 1.  It is interesting to note that most of the intensity peaks, with the
exception  of SMA~7 and SMA~9, devoid the polarized intensity. 

The subarcsecond angular resolution map shows that most of the dust polarized
emission is resolved out, specially the N--S component. This suggests that this
component arises from the resolved--out core that surrounds the compact
condensations. The polarized emission that traces the NE--SW magnetic field component
is partially detected towards MM 2 and MM 1-SMA 9. Surprisingly,  the higher
angular resolution map shows polarized emission around SMA 7, which was
undetected in the lower angular resolution maps. This is probably due to the
beam smearing: with a larger beam, SMA 7 will have contributions of different
field directions, which cancel out in the  Stokes $Q$ and $U$ maps, because they
would have different signs. The magnetic field directions in the SMA 6--7 cores
appear to be oriented in the E-W direction, with the field bending to a
north--south direction South of these cores. Most of MM 2 appears unpolarized at
the present sensitivity. 

\subsubsection{Comparison with Previous BIMA Observations}

%%  SMA:	20:39:01.20 , 42:22:48.50  
%% BIMA:	20:39:00.72 , 42:22:46.70 ( -5.3 , -1.8 )
Figure~\ref{Fbima} shows the 880~$\mu$m continuum emission of the total 
intensity (Stokes $I$) and  the magnetic fields (B) segments  from the SMA
combined data as well as from BIMA obtained at 1.3~mm  by \citet{Lai03}. At
these two wavelengths the continuum emission is dominated by the dust emission
\citep{Lai03}. The SMA dust continuum map shows a  remarkably similar morphology
to the 1.3~mm BIMA map (see Figure 7 from Lai et  al. 2003), resolving clearly
MM~1 and MM~2.  The SMA detects  slightly more linearly polarized  dust emission
than BIMA. This is because the dust  emission at 880~$\mu$m is significantly
brighter than at 1.3~mm, whereas the  sensitivity and the polarization fraction
are similar at both wavelengths.  Yet, the overall pattern is also quite
similar.  The largest differences appear South of MM~1, where the B
segments in the BIMA data are oriented in the SE--NW direction, whereas the SMA
data are oriented more toward the E-W direction (see Figure~1 from Lai et al. 2003
and Figure~\ref{Fbima} from this paper).  Indeed, the average difference between
the polarization angles of the two arrays in the SE--NW  region is $\Delta
PA$=$20$\arcdeg$\pm12$\arcdeg, whereas in the rest of the region the  difference
is only $7$\arcdeg$\pm5$\arcdeg.

\begin{figure*}[]
\includegraphics[width=17.9cm]{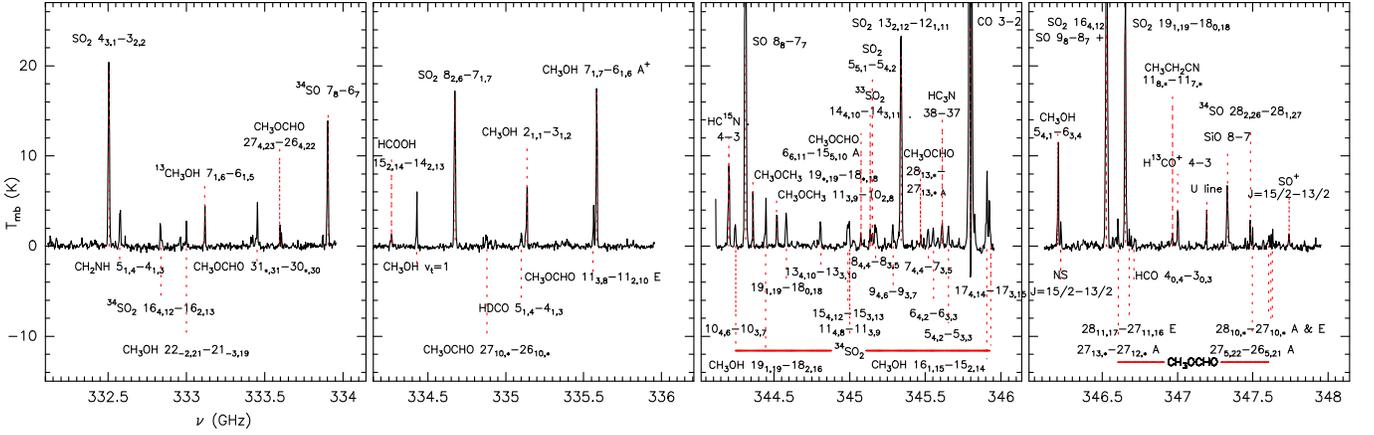}
\caption{Averaged spectra over an area of $\simeq 4$~arcsec$^2$ around SMA 6 
and SMA 7. The spectra were derived using the compact, extended and very 
extended configurations with a natural weighting
which yields a synthesized beam of  $1\farcs4\times1\farcs1$ and PA=83\arcdeg.
}
\label{Fspec}
\end{figure*}

\subsection{Molecular Lines: Dense Core Chemical Content \label{Sdense}}

\begin{deluxetable}{@{\hspace{-0.1cm}}l@{\hspace{0.3mm}}cr@{\hspace{1.9mm}}c@{\hspace{2.2mm}}l}
\tablecolumns{5}
\tablewidth{0pt}
\tablecaption{Molecular line parameters\label{T2}}
\tablehead{
\multicolumn{3}{c}{} & 
\multicolumn{1}{c}{rms\tablenotemark{a}} &
\multicolumn{1}{c}{Synthesized} 
\\
\multicolumn{1}{l}{Molecular} & 
\multicolumn{1}{c}{$\nu$} & 
\multicolumn{1}{r}{$E_{\rm L}$\tablenotemark{b}} & 
\multicolumn{1}{c}{{\hspace{-2.0mm}}(Jy/} &
\multicolumn{1}{c}{beam} 
\\
\multicolumn{1}{l}{Transition} & 
\multicolumn{1}{c}{(GHz)} & 
\multicolumn{1}{r}{(K)} & 
\multicolumn{1}{c}{{\hspace{-1.0mm}}Beam)} &
\multicolumn{1}{c}{{\hspace{-2.0mm}}FWHM ($''$), PA} 
}
\startdata
CO  3--2					&  345.796	& 17	
		& 0.10 & 0.91$\times$0.66, 89\arcdeg \\
\hcqn\ 4--3 					& 344.200	& 25	
		& 0.09 &	 1.79$\times$1.64, 62\arcdeg \\
\htcop\ 4--3					& 346.998 	& 25	
		& 0.12 & 1.80$\times$1.63, 61\arcdeg \\
\DE\ 11$_{3,9}$--10$_{2,8}$	& 344.358 & 56	
		& 0.09 & 0.90$\times$0.67, $-88$\arcdeg \\
CH$_3$CH$_2$CN 25$_{8,17}$--25$_{7,18}$ & 333.120 & 194
		& 0.08 & 0.89$\times$0.67, 89\arcdeg \\
\meta\ 18$_{2,16}$--17$_{3,14}$	& 344.109 & 403
		& 0.09 & 0.90$\times$0.67, $-88$\arcdeg \\
\enddata
\tablenotetext{a}{$rms$ noise value obtained at a spectral resolution
of 0.7~\kms.}
\tablenotetext{b}{Energy level of the lowest rotational level.}
\end{deluxetable}%

The SMA observations sample a total of  7.8 GHz bandwidth at a spectral
resolution  of 0.6\kms. They,  therefore, capture many lines in the 880~$\mu$m
band.  The  emission of most of these lines appears to be compact and mostly
associated with  SMA~6, SMA~7 and to a smaller extent with SMA~4.  The spectra
toward SMA~6  and SMA~7 (see Figure~\ref{Fspec}) clearly show that they are
dominated by  hot--core line  tracers such as methanol (\meta), sulfur monoxide 
and dioxide (SO and \sod) and methyl formate (CH$_3$OCHO).  There are also 
other hot-core tracers: oxygen--bearing species, such as dimethyl ether 
(CH$_3$OCH$_3$) and formic acid (HCOOH);  nitrogen-bearing species, such as 
cyanoacetylene (HC$_3$N), nitrogen monosulfide  (NS), methanimine (CH$_2$NH) 
and  ethyl cyanide (CH$_3$CH$_2$CN) .   The ``exotic'' sulfur monoxide ion
(SO$^+$)  is also  detected, which is a diagnostic of dissociative shock
chemistry \citep{Turner92}.  There is an unidentified line  at 347.191~GHz,
which has been previously reported toward Orion  KL/IRc2   \citep{Jewell89}. 
There are other molecular species that exhibit more extended emission, tracing
either the whole  \dr\ dense core  (\hcqn, \htcop) or the powerful outflow (CO,
SiO).  A more detailed   study of the complete molecular content detected with
this set of observations   will be reported in a forthcoming paper.  In this
paper, we focus on a selected set  of lines in order to better understand the
kinematic characteristics of the  core, the SMA~6 and SMA~7 condensations and of
the outflows, which in addition are useful  to better understand the complex
magnetic field configurations.  Table~\ref{T2} shows the list of molecules used
in this paper, with the  selected transition, rest frequency, lower energy
level, $rms$ noise level and  the angular resolution.

\subsection{Molecular Lines: Tracing the Gas Kinematics\label{Skin}}

\begin{figure}[h]
\includegraphics[width=\columnwidth]{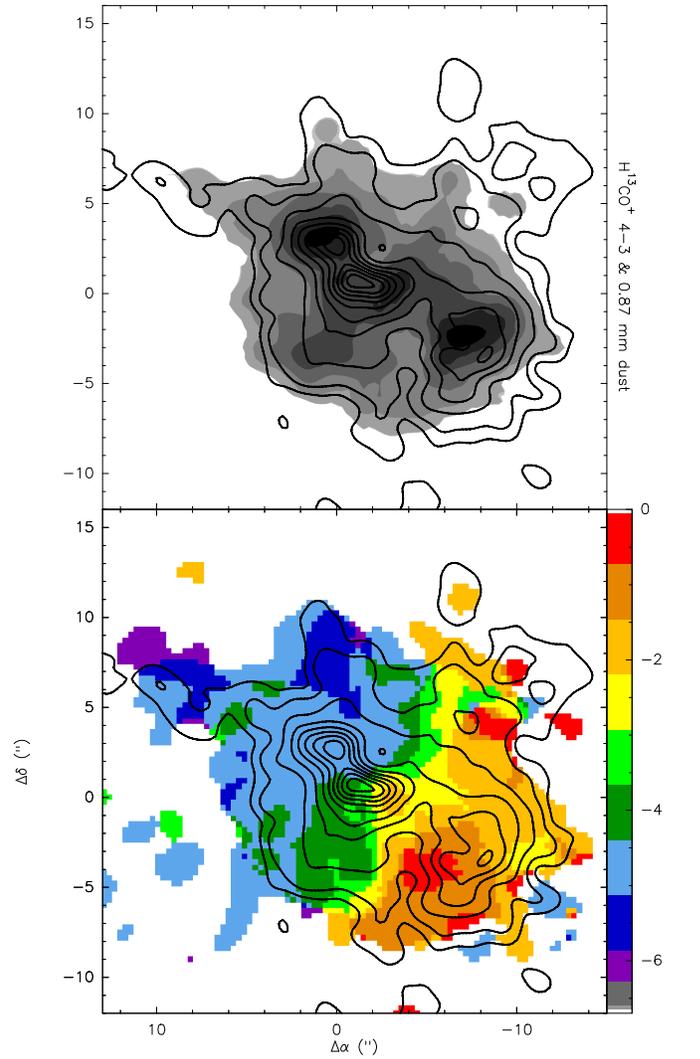}
\caption{
{\it Top panel:} gray scale image of the \htcop\ 4--3 integrated emission
overlapped with the contour map of the dust emission at an angular 
resolution of $1\farcs3$. 
{\it Bottom panel:} color image of the \htcop\ 4--3 first--order moment 
(intensity weighted mean \vlsr) overlapped with the contour map of the dust 
emission. The wedge units are in \kms.
}
\label{Fh13co+}
\end{figure}

Figure~\ref{Fh13co+} shows the integrated emission of the \htcop\ 4--3 line, as
well as its first order moment (the velocity field) overlapped with the dust 
emission at a similar angular resolution. \htcop\ is about $\sim90$ times less 
abundant than H$^{12}$CO$^+$, and it is optically thin \citep{Hezareh10}.  The 
4--3 line has a critical density of $\sim 10^7$~\cmt, which is similar to the 
averaged density found from the dust continuum observations (see 
Section~\ref{Sdust2}). All these characteristics indicate that this 
molecular  transition is, thus, a good tracer of the very dense, warm core.
Indeed, the  integrated emission appears to trace remarkably well the dust
emission. The main  difference between the dust and \htcop\ emission is toward
the condensations SMA 6 and SMA 7,  where the \htcop\ emission does not show 
a peak as the dust emission. This  suggests that this line does not trace the hot 
core--like condensations. 

\begin{figure}[h]
\includegraphics[width=\columnwidth]{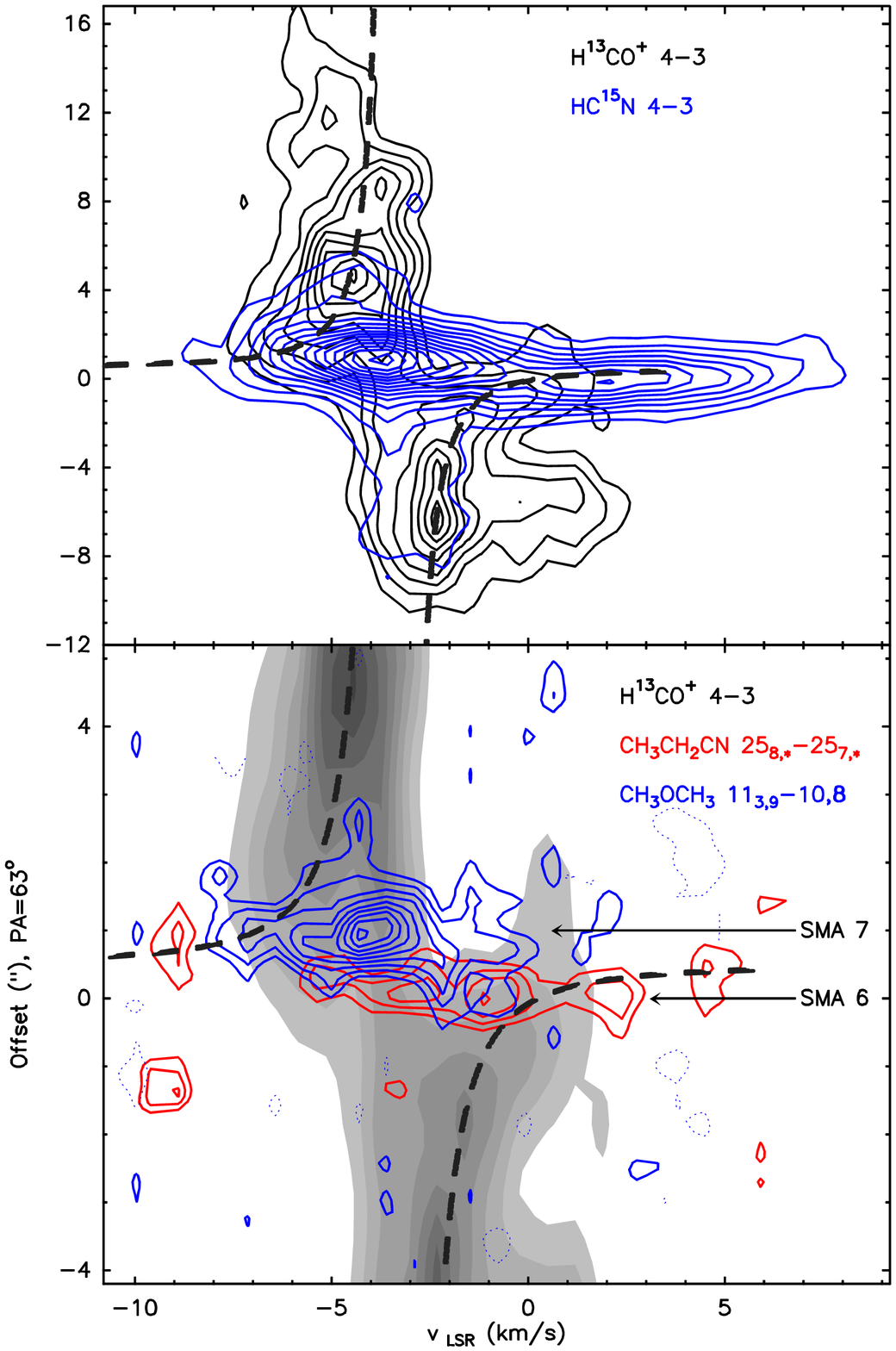}
\caption{Plot of the velocity vs. position map for a position angle of
$PA=63\arcdeg$ along the  major axis of the \dr\ dense core. The 0$''$ position
is at the location of the SMA~6 hot core  (RA$=20^{\rm h} 39^{\rm m} 1\fs00$ of
and DEC$=42\arcdeg22' 48\farcs93$). {\it Top panel:} \htcop\ (black solid lines)
and \hcqn\ (blue solid lies) lines. {\it Bottom panel:} \htcop (gray scale),
dimethyl ether (blue contours)  and ethyl cyanide (red contours). The specific
transitions are shown in the top right corner of each panel. The dashed thick
line shows the expected Keplerian rotation. 
}
\label{Fpv}
\end{figure}

The flux-weighted velocity map of the  \htcop\ 4--3 line shows a clear velocity 
gradient along the NE--SW direction, which roughly coincides with the major 
axis of the dense envelope and one of the magnetic field dominant directions. 
This velocity pattern agrees with the velocity pattern in the filament around
the core, as traced by the lower density lines \htcop\ 1--0 and N$_2$H$^+$
1--0 \citep{Schneider10}.  A
cut along the major axis of the \dr\ dense  envelope (PA=$63$\arcdeg; see the top
panel of Figure~\ref{Fpv}) shows clearly this  velocity gradient. The eastern side
of the envelope is blue-shifted with respect to the western side, with systemic
velocities of \vlsr$\simeq -4.7$~\kms\ and \vlsr$\simeq -2.3$~\kms,
respectively.   Using as a reference the distance between the eastern and
western edges of the envelope along the major axis (\ie\ along the NE--SW
direction),  $\simeq 20''$,  and the observed velocity difference between these
two edges, $\simeq 2.4$~\kms,  we estimate a velocity gradient over the whole 
core of $\simeq 18$~\kms\ pc$^{-1}$.  At about $4''$ (5600 AU in projection) 
from SMA~6 the gas velocity starts to increase in a Keplerian--like motion (\ie\
the  blueshifted/redshifted gas becomes bluer/redder  toward the center). However,
the  lack of \htcop\ 4--3 emission associated with the SMA 6--7 condensations
does not allow us to  inspect the kinematics at the  very center.

\begin{figure}[h]
\includegraphics[width=7.8cm]{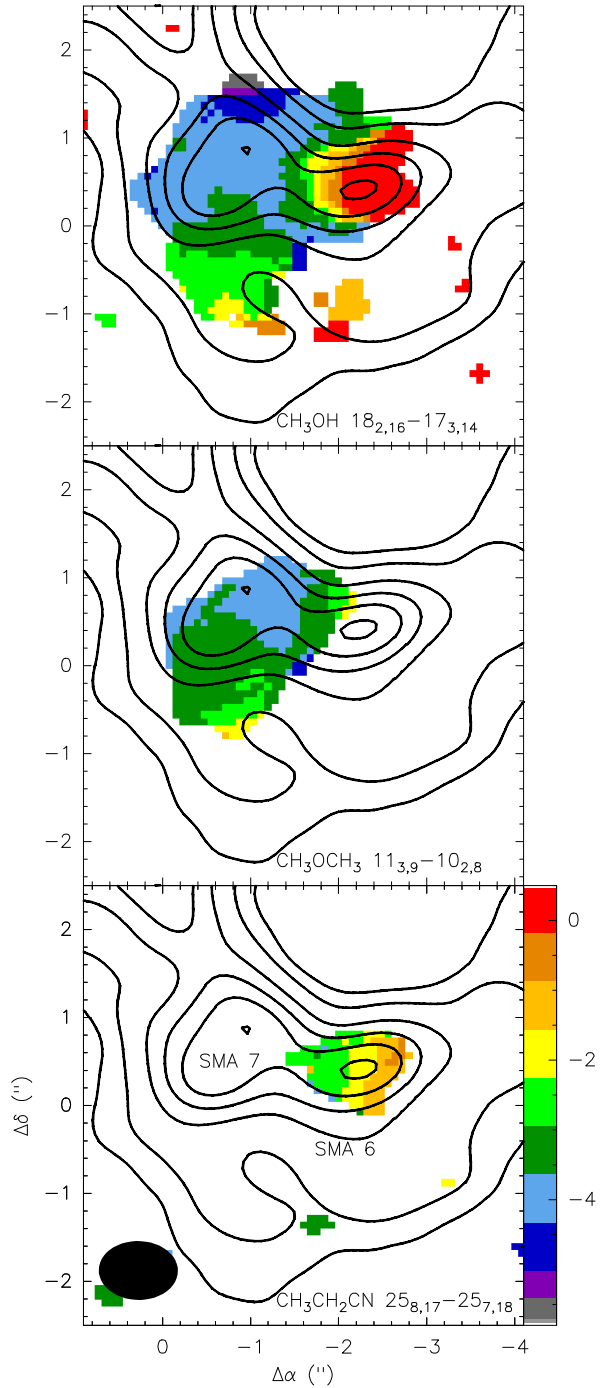}
\caption{Contour maps of the sub-arcsecond angular resolution dust emission
overlapped with the  color images of the  first-order  moment maps of three
selected molecular transitions.  The wedge units are in \kms.
{\it Top panel: } \meta\ 18$_{2,16}$--17$_{3,14}$;
{\it Middle panel: } \DE\ 11$_{3,9}$-10$_{2,8}$;
{\it Bottom panel: } CH$_3$CH$_2$CN 25$_{8,17}$--25$_{7,18}$.
}
\label{Fmol}
\end{figure}

In order to check the kinematics at the center of the core, we have made  maps
of four different molecular lines (see Table~\ref{T2}). The HC$^{15}$N  4--3
emission is similar to \htcop\ 4--3 but slightly more compact and brighter 
toward SMA~6--7.  The position-velocity cut of the \hcqn\ line along the major
axis  is similar to the \htcop\ line at scales of few arcseconds. However, there
are significant differences between these two tracers at distances less than 
$\la 1\farcs5$ ($\la 2000$~AU) from the center. The  \hcqn\ line is much
brighter and  its emission arises from two components that are not traced by the
\htcop line: one at  $-4$~\kms, associated with SMA~7, and another one at
2~\kms\ arising from SMA~6.  These two components appear to break  the
Keplerian--like kinematic behavior  observed in the \htcop\ line. To further 
investigate the kinematics around the densest  part of the \dr\ center, SMA~6 
and SMA~7, Figure~\ref{Fmol} shows the first-order moment maps of three hot-core 
molecular transitions, overlapped with the dust continuum maps. All of them  are
obtained at a sub-arcsecond angular resolution.  The \meta\ line -- which  has a
high excitation energy level -- traces very well the two hot cores SMA~6
and  SMA~7.  The velocity behavior resembles well the one from  the \hcqn\ line,
particularly the red-shifted component at 2~\kms\ associated with SMA~6.  The
dimethyl ether line emission traces only SMA~7, which is observed in both
\meta\ and \hcqn\  lines with a systemic velocity of $-4$~\kms. This line shows
a velocity gradient of $\simeq 1$~\kms\ roughly along the SMA~7 major axis,
\ie\ along the NW-SE direction. On the other hand, the  ethyl cyanide line
traces only the dense gas associated with SMA~6. It also shows a small velocity
gradient of 1~\kms but along the E-W direction. Interestingly, the line peaks
at a \vlsr\ velocity of  $\simeq -2$~\kms. This is quite different from the
\meta\  and the \hcqn\ line  emission around SMA~6 (see the top panel of
Figure~\ref{Fpv}).  The bottom panel of Figure~\ref{Fpv} shows the position-velocity
cuts  for the dimethyl ether and ethyl cyanide overlapped with  the \htcop. As
already shown by \hcqn, the gas traced by these two hot-core lines apparently
does  not follow the Keplerian--like behavior of the \htcop\ emission. In
Section~\ref{dis} we discuss the interpretation of these differences.

% Moment maps of selected dense tracer

\subsection{CO 3--2: A Molecular Outflow Tracer}\label{co}

\begin{figure}[h]
\includegraphics[width=\columnwidth]{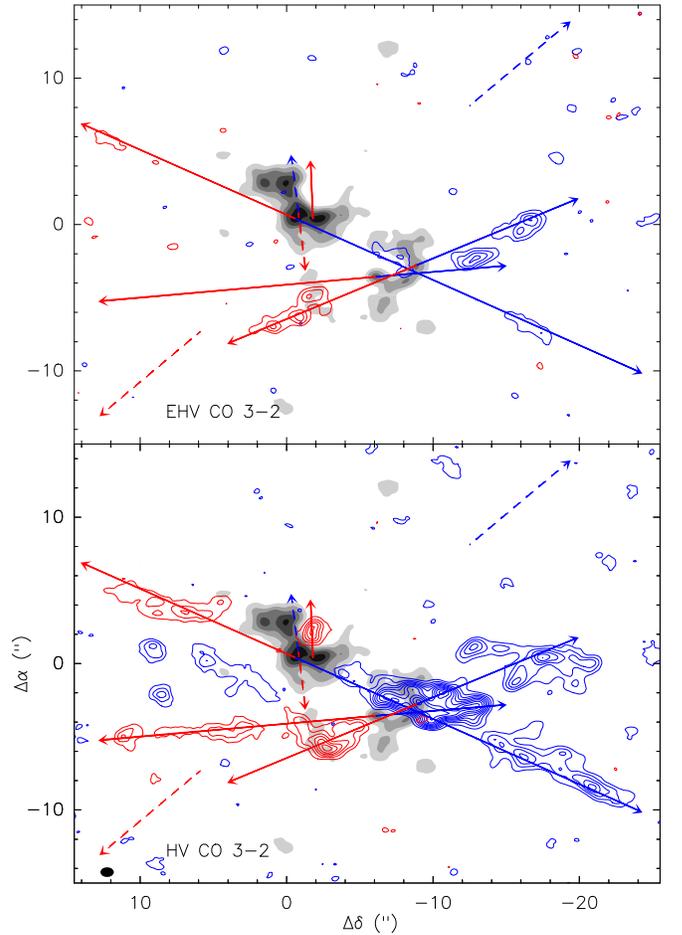}
\caption{Maps of the red-shifted (red contours) and blue-shifted (blue contours)
emission of the CO 3--2 line.  
{\em Top panel:}  EHV component obtained by averaging the  emission over
50~\kms\ centered  at a velocity $\pm 65$~\kms\ with respect to  the system
velocity of the \dr\ core, \vlsr\ $\simeq -3$~\kms.  
{\em Bottom panel:}  HV component obtained by averaging  the emission over
20~\kms\ centered at a velocity $\pm30$~\kms\ with respect  to the system
velocity. The gray-scale map shows the dust continuum image  of the high-angular
resolution. The synthesized beam of the CO maps is shown  in the bottom left
corner of the bottom panel. Solid arrows show the observed  outflows in the CO
3--2 maps. Dashed arrows show the bipolar outflows detected previously by
\citet{Minh11} from the SMA H$_2$CS observations (compact N--S  arrows centered
around SMA 6--7) and \citet{Vallee06} from single-dish  observations of CO 3--2
(SE--NW arrows).
}
\label{Fco}
\end{figure}

Figure~\ref{Fco} shows the high-velocity (HV) emission of the CO 3--2  line at an 
angular resolution of $\simeq 1''$ for two different velocity ranges in the
blue  and red lobes. The overall pattern agrees with the previous lower angular 
resolution BIMA interferometric  maps ($\simeq 5''$)  of the CO 2--1 line by  
\citet{Lai03}. The CO 2--1 maps  showed two bipolar outflows oriented roughly 
in the E--W direction with  the red-shifted lobes in the eastern part. The SMA 
CO 3--2 maps show that most of the emission is distributed similarly to the  CO
2-1 emission, but the higher angular resolution reveals a more complicated 
morphology.  The extremely high--velocity (EHV) CO emission (velocities from 40
to 90~\kms\  with respect to the cloud velocity) appears to arise from a 
bipolar structure with a position angle in the direction of the red-shifted
lobe  of about 110\arcdeg. The origin of the outflow appears to be in MM~2,
possibly  from SMA~3 or SMA~4.  At HVs (velocities from 20 to
40~\kms\ with  respect to the cloud velocity) there are two highly collimated
bipolar outflows  with position angles of 95\arcdeg and 65\arcdeg. The first one
appears to also  arise from MM~2, possibly SMA~4, although we cannot discard
source SMA~3.  It is  possible that this HV emission  is part of the
same outflow as the  EHV bipolar component, as suggested
from methanol and  formaldehyde observations \citep{Zapata12}. The second
outflow arises from MM~1,  possibly from  SMA~6 or SMA~7. There is an isolated
redshifted clump only $2''$  away of SMA 6 and 7, without a blue-shifted
counterpart in the same velocity range.  It is very close to a compact, low
velocity outflow detected in H$_2$CS \citep{Minh11}.

Interestingly, the interferometric maps of the CO 3--2 outflows are strikingly 
different from the single-dish maps of the same transition \citep{Vallee06}. 
These lower angular-resolution ($\sim 14''$) maps show a low-velocity  outflow
with a position angle of roughly 130\arcdeg, with the blue and red  lobes
located NW and SW, respectively, of the \dr\ center.  This low-velocity outflow
is not seen in the SMA maps because it is probably too extended, and therefore 
most of the emission is filtered out.

% Moment maps

\section{Analysis: Statistical derivation of the magnetic field
strength}\label{ana}

Figure~\ref{Fpol} shows clearly that at the scales traced by the SMA, the 
magnetic field segments in the \dr\ region do not follow a defined homogeneous 
pattern as, e.g., the hourglass shape reported in some low- and high-mass 
star-forming cores \citep{Girart06,Girart09}. However, if we take into account 
the large-scale polarization maps, the magnetic field segments show significant 
coherence in all the maps except the one tracing the densest regions. No simple 
analytical models are available to be compared with this complex magnetic field
and mass distribution  \citep[see, e.g.,][]{frau11}. Therefore, in order to
extract  physical information, a statistical approach seems to be the best
available  option. 

\begin{figure}
\centering
\includegraphics[width=\columnwidth]{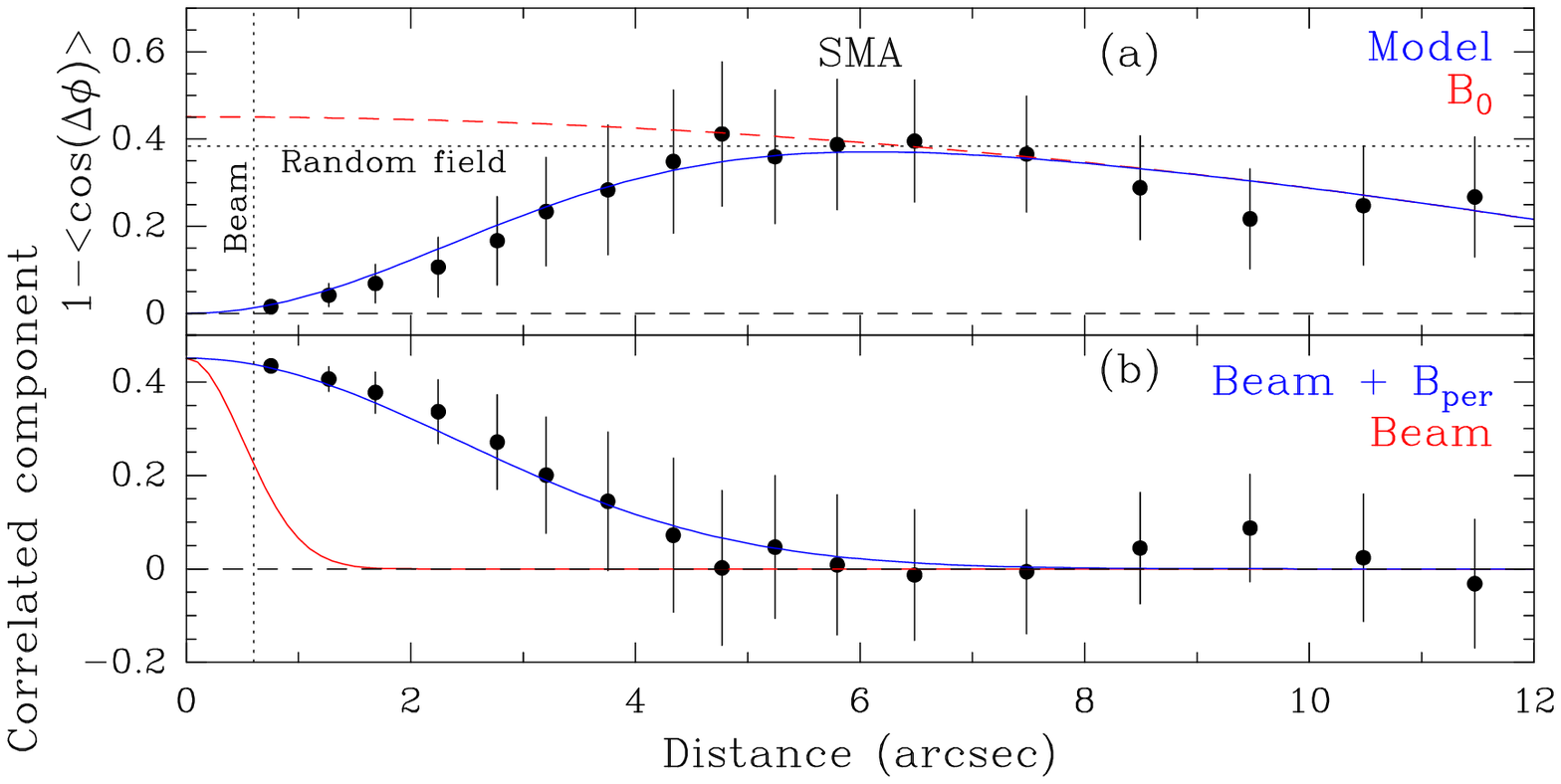}
\includegraphics[width=\columnwidth]{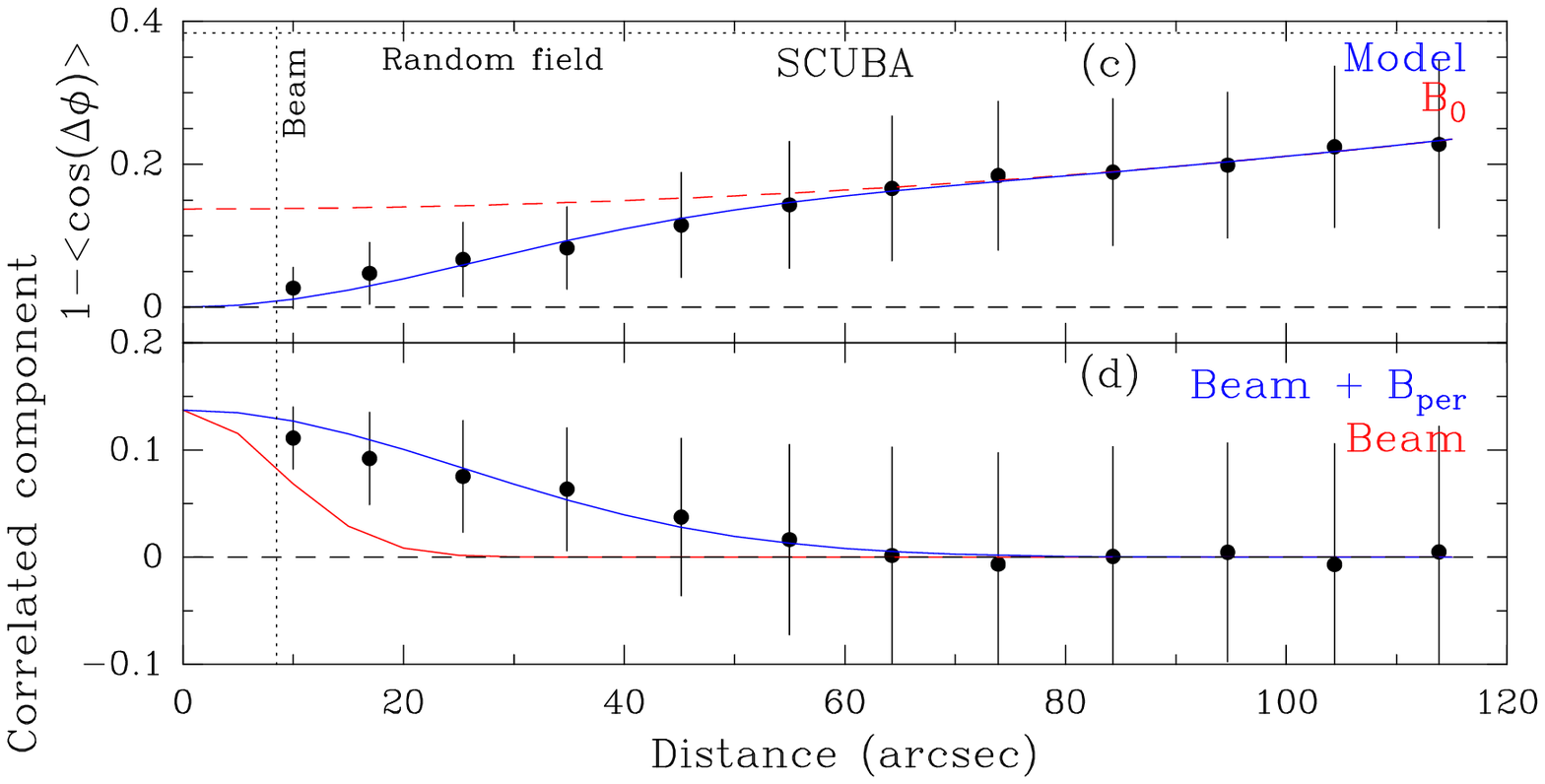}
\caption{
{\it Panels (a) and (c):}  angular dispersion function  using the magnetic  field
segments detected towards \dr, obtained using a Nyquist sampling. The data points 
(shown as dots) and the error bars are the mean and standard deviation of all the 
pairs contained in
each bin.  The red dashed line  shows the fitted $f_{{\rm NC}}(0)\, + a'_{2}\,l^{2}$. 
The dotted vertical line gives the FWHM of the  synthesized beam. The dotted 
horizontal line shows the expected value for a random magnetic field. The blue 
line shows the best fit to the data (Equation (\ref{eq1})). 
{\it Panels (b) and (d):} the dots represent the correlated component of the best fit
to the data. The dashed line marks the zero value. The solid red line shows the
correlation due to the beam, and  the blue line  shows the correlation due to
the beam and the turbulent component of the magnetic field.
Panels  (a) and (b) show the data from SMA. Panels (c) and (d) show the data from SCUBA.
}
\label{fig-houde09}
\end{figure}

To statistically analyze the data we have estimated the angular dispersion 
function, 1--$<$cos$[\Delta \Phi (l)]$$>$, where $\Delta \Phi (l)$ is the 
difference between the polarization angles measured for all pairs of points 
separated by a distance $l$.  Note that for small values of $\Delta \Phi (l)$,  
1--$<$cos$[\Delta \Phi (l)]$$> \simeq 1/2$$<$$\Delta \Phi^2(l)$$>$, which is 
the  second--order structure function of the polarization angles. This function 
gives information on the behavior of the dispersion of the polarization  angles
as a function of the length scale in the dense molecular gas
\citep{hildebrand09, houde09, houde11, franco10, koch10}. 
Figure~\ref{fig-houde09}  shows this function applied to the SMA and SCUPOL
polarimetric data (the top and bottom panels, respectively).  For the two 
telescopes, due to the effect of the limited angular resolution, the angular 
dispersion function is zero at $l=0$ and then smoothly increases with the 
length scale.  However, the overall behavior is quite different between the  two
telescopes. On one hand,  in the SMA polarization data the second order 
dispersion function increases with angular separation $l$, reaching  values 
compatible with a random magnetic field  
\citep[$\simeq$52\arcdeg:][]{poidevin10} at scales of $4''$  (5600~AU). This
behavior is similar to the one found in NGC~7538~IRS~1  \citep{frau13}. 
Interestingly at scales larger than $8''$  ($10^4$ÊAU), the angular dispersion
starts to decrease to values of  $\simeq 0.2$ (this is equivalent to an angular
dispersion of $\sim 35$\arcdeg).  On the other hand, the SCUPOL data, which
traces much larger scales than the SMA ($\ga 20''$), shows that the dispersion
never reaches the value  expected for a random field. Indeed, the maximum
angular dispersion at $l=100''$ ($10^5$ÊAU) is roughly 0.2.

%______________________________________________

%________________________________________________________________
\begin{table}
\caption{Angular Dispersion Function Fit Parameters\tablenotemark{a}}\label{tab-houde09}
\begin{center}
\begin{tabular}{ccc}
\hline
Parameter & SMA & SCUPOL \\
\hline
$\delta$ (mpc)		& $16.9\pm1.6$		& $151\pm21$ \\
$f_{{\rm NC}}(0)$ 	& $0.45\pm0.04$	& $0.137\pm0.015$ \\
$\Delta'$ (pc)		& 
			$\simeq 0.08$	& $0.34$\tablenotemark{b} \\
$n({\rm H_2})$ (cm$^{-3}$) &
			$1.0\times10^7$ 	&
			$2.0\times10^5$\tablenotemark{c} \\
$\delta V$ (km~s$^{-1}$) &
			1.0			& 0.8\tablenotemark{d} \\
$\langle B^2_{\rm t} \rangle / \langle B^2_0 \rangle$ & 
			$\simeq 0.92$		& $\simeq 0.16$  \\
$\langle B\rangle_{pos}$ (mG) & 
			$\simeq 2.1$ 		& $\simeq 0.62$ \\
$\langle B\rangle$ (mG) & 
			$\simeq 2.1$	 	& $\simeq 0.9$4 \\
\hline
\tablenotetext{a}{Following \citet{houde09}. See Section~\ref{ana} for the 
definition of the different parameters.}
\tablenotetext{b}{Assumed thickness for the SMA (see Section~\ref{ana}). For the 
SCUPOL data we adopt the value derived by \citet{Hennemann12}.}
\tablenotetext{c}{
Value derived from single-dish 1.3~mm dust continuum
observations \citep{Motte07}.}
\tablenotetext{d}{Value derived from single-dish \htcop\ 1-0 observations
\citep{Schneider10}.}
\end{tabular}
\end{center}
\end{table}
\vspace{0.5cm}
%__________________________________________________________________

Assuming a stationary, homogeneous, and isotropic magnetic field strength and  a
magnetic field turbulent correlation length, $\delta$, smaller than the 
thickness of the cloud $\Delta'$, \citet{houde09} have shown that the angular 
dispersion function can be used to estimate the importance of the magnetic
field. We have used Equation (42) from \citet{houde09}, which takes into account
the  smearing effect of the beam and the line-of-sight integration, to estimate
the  importance of the field. Under these assumptions, the dispersion function
can be  rewritten as
\begin{equation}\label{eq1}
1-\langle \cos\left[\Delta\Phi\left(\,l\,\right)\right] \rangle
\simeq
f_{{\rm NC}}(0)\, 
\left[1-e^{-l^2/2(\delta^2+2W^2)}\right]
\\
+\sum^{\infty}_{j=1}a'_{2 j}\,l^{2 j} ,
\label{eq-houde09}
\end{equation}
where $l$ is the length scale and $W$ is the beam ``radius''\footnote{ $W = {\rm
FWHM}/\sqrt{8 \, ln 2}$, where FWHM is the full-width at half-maximum of the
beam.}. The summation on the right hand side of the equation is the contribution
from the ordered component of the  magnetic field, and $ f_{{\rm NC}}(0)$ is the
value of the  correlated component at the origin (shown in the bottom panel of
Figure~\ref{fig-houde09}).   This value depends on the energy ratio between the
turbulent or perturbed  magnetic field and the ordered large-scale magnetic
field,   $\langle B^2_{\rm t} \rangle / \langle B^2_0 \rangle$,  and the number 
of independent turbulent cells contained in the column of dust probed 
observationally, $N$:   
$f_{{\rm NC}}(0)=(\langle B^2_{\rm t} \rangle / \langle B^2_0 \rangle) \,(1/N)$.  
According to \citet{houde09}, 
\begin{equation}
N = \Delta ' \, (\delta^2 + 2 W^2) / (\sqrt{2 \pi} \delta^3)
\label{eq-N}
\end{equation}
where $\Delta '$ is in the effective thickness of the molecular cloud, which
is expected to be somewhat smaller than the cloud thickness.

\subsection{SMA Polarization Data}

In the case of the SMA polarization data, Figure~\ref{fig-houde09} shows that  at
scales larger than $l \sim 4''$, the magnetic field has statistically values 
similar to what is expected for a random field (though for $l \ga 8''$ the
dispersion  function decreases below the random field value).  However, this
does not imply  that the magnetic field is random (see Section~\ref{dis} for a
discussion on this  issue).  The best fits to the SMA \dr\  polarimetric data
(see Table~\ref{tab-houde09} and the top two panels of  Figure~\ref{fig-houde09})
lead to a turbulent magnetic field correlation length  of $\delta=
2\farcs33\pm0\farcs22$ ($16.9\pm1.6$~mpc at 1.5~kpc).    The derived value of
the correlated component at the origin is   $f_{{\rm NC}}(0)$ $\simeq 0.45$. A
reasonable approximation is to assume that the core's effective thickness,
$\Delta '$  is similar to the average diameter of the dense core measured in the
plane  of the sky with the SMA \citep{koch10}, which in our case is $\simeq
10\farcs4$ ($\sim 1.6 \times 10^4$~AU, see Figure~\ref{Fpol}). Following
Equation~(\ref{eq-N}), this  yields to  $N \simeq 2$ turbulent cells along the
line-of-sight. This implies that  $\langle B^2_{\rm t} \rangle / \langle B^2_0
\rangle \simeq 1$, \ie\ there is equipartition between the perturbed and ordered
magnetic field energies.

The Chandrasekhar--Fermi (CF) equation can be used to derive the magnetic 
field strength in the plane of the sky, 
$\langle B^2_0\rangle^{1/2}\propto \delta V \, n({\rm H_2})^{1/2} 
\left[ \langle B^2_{\rm t} \rangle / \langle B^2_0 \rangle\right]^{-1/2}$
\citep[e.g., see Equation (57) by][]{houde09}. Table~\ref{tab-houde09} shows the values
used for the velocity dispersion, $\delta V$, and for the volume  density,
$n({\rm H_2})$.  We estimated the velocity dispersion from the \htcop\ 4--3
data, since, as shown in section~\ref{Sdense}, its emission  is well correlated
with the 880~$\mu$m dust emission. For the volume  density, we used the value
derived in Section~\ref{Sdust2}. From the combination of both results, the ordered
large-scale magnetic field strength component in the plane-of-sky, 
$\langle B^2_0 \rangle^{1/2}$, $\simeq 2.1$~mG.

\subsection{JCMT-SCUPOL Polarization Data}

For the SCUPOL, we follow the same steps of the previous subsection. However, we
only compute the statistics for length scales  less than $2\farcm0$, since we
want to fit the scales in  Figure~\ref{fig-houde09} where the ordered,
large-scale component  is  approximately linear.  The physical parameters
derived from the analysis  are shown in the right column of
Table~\ref{tab-houde09}.  The turbulent  correlation length is about 0.15 pc
which is significantly larger than for the SMA. The derived value of $f_{{\rm
NC}}(0)$ is  $\simeq 0.14$.  For the effective  thickness we adopt the value of
the filament width obtained from {\it Herschel} observations, 0.34~pc
\citep{Hennemann12}. Using the velocity dispersion  and the volume density
values reported in the literature for the filament, we  derive a
perturbed-to-ordered magnetic energy ratio significantly lower than the value
for the \dr\ core, $\simeq 0.2$. The plane-of-the-sky magnetic  field strength
is 0.62~mG, similar to the value derived  previously \citep{Vallee06}.

\section{Discussion: The relevance of the magnetic fields}\label{dis}

\subsection{Comments on Individual Sources}

It is noteworthy that most of the dust peaks of the different condensations 
(Figure~\ref{Fpol}) are devoid of polarized emission.  This can be due to beam 
cancellation at the center of the different condensations,  where gravity 
pulls the field lines to the  center \citep{frau11}. Of the different submillimeter
condensations, only three, namely SMA~4, SMA~6 and SMA~7 show 
clear signs of on--going  star formation.

SMA~4 has a very compact dust distribution, and it has associated emission from 
shock-excited dense tracers. In addition, it appears to be the powering
source of the east-west highly collimated CO outflow. The strong methanol and 
formaldehyde emission associated with this outflow \citep{Zapata12} suggests
that this outflow is strongly interacting with the dense gas.  

With a mass of $\simeq 23$~\msun\ \citep{Zapata12}, SMA~6 is one of the most 
massive condensations embedded in the \dr\ core near the geometrical center of the 
core.  A very compact and dense molecular outflow has been detected in the  N--S
direction that appears to be centered on this source \citep{Minh11}.  The  SMA
observations show that it has a hot--core like chemistry. The ethyl  cyanide, a
hot-core tracer, is present only in this source (the bottom panel of 
Figure~\ref{Fmol}).  This tracer shows a clear velocity gradient along the 
east--west direction, \ie\ perpendicular to the associated compact outflow.
Thus,  this velocity gradient probably indicates  rotation.  The red--shifted
component  seen in methanol and \hcqn\ (Figures \ref{Fpv} and \ref{Fmol}) is
likely tracing  shock--excited emission from the compact outflow.  

SMA~7 is the other massive condensation with a mass similar to SMA~6. It 
also has a hot--core chemistry, although it is different with respect to SMA~6 
(J. M. Girart, private communication). For example, dimethyl ether is only detected in 
this source (see the middle panel of Figure~\ref{Fmol}). Its emission appears to be 
extended in the NW--SE direction, with a velocity gradient along the same 
direction. This velocity gradient is roughly perpendicular to the highly collimated 
CO outflow with  PA=65\arcdeg. This suggests that SMA~7 is the powering 
source of this outflow. 
Both SMA~6 and SMA~7 have velocity gradients that does not match the large scale 
velocity gradient seen in the core through the \htcop\ 4--3 emission.  Recent simulations of non-idealized magnetized massive cores show that turbulence can generate the observed misalignment \citep{Seifried12}.

\subsection{Interpretation of the Statistical Analysis}

The polarization angle dispersion shows relatively high values for the SMA 
observations, as high as those expected for a random field. However, this does 
not imply that there is a lack of an ordered field.  As an example,  the
classical  ordered hourglass magnetic field expected in a magnetized core with
little  turbulence and rotation -- which has been observed in some protostars 
\citep{Girart99, Girart06, Lai02, Alves11} -- will have a radial pattern in the 
plane of the sky  in the case of a face--on configuration \citep{frau11,  
Padovani12, Kataoka12}. 
A similar case would be a toroidal field (due to rotation) also face--on.
Both patterns would also appear in the structure function  with values close to 
the  ones expected for a random field.  For a qualitative assessment, we have 
computed the second-order structure function on the simulations shown in the 
two bottom panels of Figure 8 by \citet{Padovani12}. These two panels show the 
B segments of a toroidal magnetic configuration seen face--on at two different 
times of the collapse of a magnetized core using the RAMSES code \citep{Fromang06, 
Hennebelle08, Hennebelle09}.  Despite that the conditions are different 
(the simulations used are for a low mass star forming core), we found that the 
overall statistical trend of the toroidal field simulations confirm that the 
second-order structure function behavior observed in the SMA data can be explained 
simply by a toroidal field (see Section~5.6) rather than a very turbulent medium.  

The statistical analysis carried out with the SMA polarimetric data toward \dr\ 
yields values of the  turbulent length scale, $\simeq 17$~mpc,  and of the
magnetic field strength component in the plane of the sky, $\simeq 2.1$~mG, that
is in very good agreement with  the values  derived from a completely
independent method by \citet{Hezareh10} who  found 9~mpc and 1.7~mG. They 
computed these values from the correlation of  the velocity dispersion of the 
coexisting neutral and ionized species H$^{13}$CN  and \htcop, using their
rotational   4--3 line. Note that as stated previously, our  SMA data show that
the \htcop\ 4--3  correlates well with the 880~$\mu$m dust  emission, \ie\ they
trace the same gas.  This good agreement gives confidence in the values derived
despite the uncertainties  of the  analysis method. Furthermore, the value of
the correlation length derived with the SMA is also within a factor of two of the
values reported in the literature for  interferometric observations of three
other massive star  forming regions,  W51, Orion KL/Irc2 and NGC 7538 IRS1
\citep{koch10, houde11,  frau13}. 

The analysis done with SCUPOL gives a correlation length scale  significantly
larger than the value found with the SMA.  However, the  field strength (in the
plane of the sky) is similar to the value derived by \citet{Vallee06} who were
using directly the C-F method.

\subsection{Turbulence versus Magnetic Fields}

The line width in the \dr\ core is larger than in the filament 
(Table~\ref{tab-houde09}).  This is apparently in contradiction with the  Larson's 
law \citep{Larson81}.  However, in the context of very active massive star and 
cluster formation the dynamical process in dense cores, e.g., infall, rotation 
and outflows can yield a line width in the high density gas larger than the 
line width in the envelopes \citep[e.g.,][]{Zhang02, Galvan10, Keto10}. 
This is the case for DR21(OH), as shown by the observed signatures of the 
very active star  formation activity: the richness of masers where some of them 
are clearly associated with outflow  activities \citep[e.g.,][]{Kurtz04, Hakobian12};
and the molecular dense tracers showing strong emission associated with the outflow 
being powered by protostars within the  core \citep{Lai03, Zapata12}.  

Figure~\ref{fig-houde09} shows that  the angular dispersion function  has a
clearly disturbed behavior only at core scales.  Nevertheless, the strongly
perturbed,  apparently random, field appears to happen only in a small range of
scales: 6,000--12,000~AU ($\simeq 4''$--$8''$). At larger and smaller scales the
angular dispersion decreases below the random field value. Indeed, the magnetic field 
threading the parsec--scale filament appears more  ordered. Thus as in the case of 
the line width, the increase of the turbulent or disordered field in the \dr\ core 
can be a consequence of the active star formation activity. 
In any case, in spite of this large dispersion in the  core, the ordered magnetic 
fields are roughly in energy equipartition with turbulent or perturbed components 
of the field. In the filament, the ordered field dominates, energetically, over the 
turbulent/disordered  component.

\subsection{Gravitational Force versus Magnetic Fields}

A key parameter to estimate the relevance of the magnetic field with respect to
the gravitational force is the mass--to--magnetic flux ratio. This
ratio in terms of the critical value is 
$7.6 \times 10^{-21} [ N({\rm H_2}) / {\rm cm^{-2}} ] [ B / {\rm \mu G} ]^{-1}$ 
\citep{Crutcher04}. 
In order to properly use this equation, we first should estimate the total 
magnetic field strength. Fortunately,  there are Zeeman measurements of the 
line--of--sight component of the field: \citet{Crutcher99} carried out 
spectro-polarimetric observations of the CN  1--0 line around the \dr\ core. 
The Zeeman splitting was detected in two different velocity components,  $v_{\rm
LSR}=-4.7$ and $-0.9$~\kms, yielding magnetic field strengths of  $B_{\rm los} =
-0.36\pm0.10$ and $-0.71\pm0.12$~mG, respectively. The CN  observations cover a
significant part of MM 2 and trace gas at densities of $\sim 10^6$~\cmt. Recent
interferometric observations of the CN 1--0 line show that the $-4.7$~\kms\ component
is associated with the core, whereas the $-0.9$~\kms\ component arises from
widely distributed CN emission \citep{Crutcher12}. Thus,  it is reasonable to
assume $B_{\rm los}Ê\simeq 0.36 $~mG for  the whole dense core, MM 1 and MM 2,
detected with the SMA.  Therefore,  the total magnetic field strength of \dr\
is  $\simeq 2.1$~mG.  For the column density, the SMA observations yield a value
of  $N({\rm H_2}) \simeq 1.6 \times 10^{24}$~\cmd (see Section~\ref{Sdust}).  This
implies a  mass--to--magnetic flux ratio  of about 5.9 times the critical
value.  Since there is significant star formation activity, it is expected that
there is already a significant mass accreted onto the protostars embedded in
\dr. This suggests that this ratio is somewhat larger. In any case, this result
implies that the magnetic field  energy  is not enough to provide support
against gravity. Consequently, a global gravitational collapse  is expected in
the core. 

Similarly, we can estimate the mass--to--magnetic flux ratio for the large-scale
dense filament traced by the single-dish SCUPOL data. We assume that the
line--of--sight field strength of the filament is 0.71~mG. This is the value
found by \citet{Crutcher99} for the CN velocity component at $v_{\rm
LSR}=-0.9$\kms,  which is the typical systemic velocity for the whole filament
as traced by the  dense molecular tracers \citep{Schneider10}. With this $B_{\rm
los}$ value, the total magnetic field strength for the filament is  $\simeq
0.94$~mG.  Since the average column density of the filament is   $4.2 \times
10^{23}$~\cmd \citep{Hennemann12}, the mass--to--magnetic flux  ratio is 3.4.
This is lower than toward the \dr\ core, but it is still supercritical.
Therefore, it is  expected that the star formation process has already started 
along the filament. And indeed, molecular line observations reveal infall
motions  as well as the presence  of some molecular outflows along the filament 
\citep{Schneider10}.

An independent and complementary analysis of the role of the magnetic field 
is provided by the polarization -- intensity gradient method \citep{koch12a}.
In this technique, dust emission and magnetic field morphologies are interpreted
as the overall result of gravity, pressure and field forces. Magnetic field 
orientations and dust emission gradient orientations reveal a correlation where
the difference $\delta$ in their orientations can be linked to the magnetic
field strength \citep[Figure 3 in ][]{koch12a}. As a result, a local magnetic
field strength can be calculated at all positions where polarized emission 
is detected. Additionally, the method leads to an estimate of the local 
magnetic field significance relative to gravity, $\Sigma_B$, based on 
measurable angles only.

\begin{figure}
\centering
\includegraphics[width=\columnwidth]{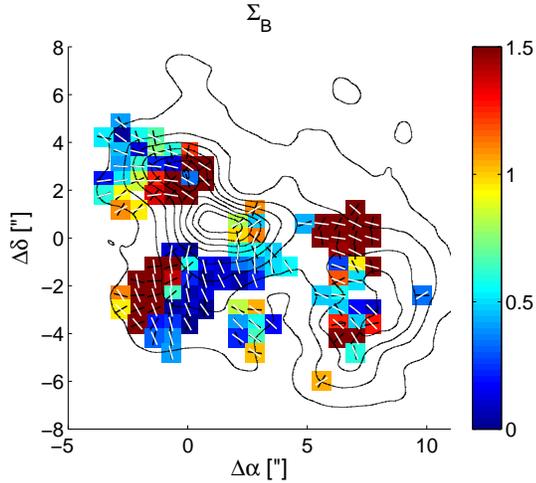}
\caption{
Map of the local field significance. The dust continuum emission is shown 
in black contours. The color wedge indicates the force ratio, $\Sigma_B$, field 
tension force over gravitational pull. Values with $\Sigma_B <1$ indicate local
gravitational collapse. White segements display the magnetic field orientations
similar to the bottom panel in  Figure~\ref{Fpol}. The dust emission gradient
orientations are shown with black segments at the positions of polarized
emission. The deviation between the two orientations is the angle 
$|\delta| \le 90^{\circ}$, with a map-averaged 
$\langle|\delta|\rangle \approx 40^{\circ}$. A clear correlation in the 
two orientations is apparent with $\mathcal{C}=0.74$. 
Areas in blue translate into a mass-to-flux ratio of about 2-3 times the
critical value.
}
\label{fig-sigma_B}
\end{figure}

We have applied this technique to the SMA maps shown in the bottom panel in 
Figure~\ref{Fpol}. The force-ratio map, $\Sigma_B=F_B/F_G$ where $F_B$ is  the
magnetic field tension force and $F_G$ is the gravitational pull, is shown in
Figure~\ref{fig-sigma_B}. The derivation of $\Sigma_B=\sin\psi/\sin\alpha$ makes
use of $\delta=\pi/2-\alpha$ in combination with an additional angle $\psi$
between the dust emission gradient and the local gravity direction.  The
map-averaged deviation is $\langle|\delta|\rangle \approx 40^{\circ}$ with a
standard deviation of $26^{\circ}$, and a correlation coefficient
$\mathcal{C}=0.74$ for the intensity gradient -- field alignment. These values
are similar to the ones found for other cores, as e.g. in \citet{Tang13}. The
average force ratio after removing some outliers is 
$\langle\Sigma_B\rangle\approx 0.8$. This indicates that on average the magnetic
field is overwhelmed by gravity, and thus, a gravitational collapse is  enabled
($\Sigma_B < 1$). The local force ratio can furthermore be  transformed into a
local mass-to-flux ratio \citep{koch12b},  $M/\Phi \propto
(\sin\psi/\sin\alpha)^{-1/2}$. For the blue-colored patches in 
Figure~\ref{fig-sigma_B}, this leads typically to mass-to-flux ratios of about  2
to 3 times the critical value. This supports the above finding of a globally
supercritical core based on measured values for $ N({\rm H_2})$ and $B$. We,
nevertheless, also acknowledge that some isolated patches (in red in 
Figure~\ref{fig-sigma_B}) point to a locally dominating role of the field where
the  magnetic field tension still outweighs gravity.

%______________________________________________

\subsection{Magnetic Field Flux Diffusion at Core Scales}

The magnetic field strength in the \dr\ core is only a factor two higher than the
averaged value in the parsec scale filament, whereas the volume density is more
than one order of magnitude higher. This suggests  magnetic flux diffusion or
dissipation.  Assuming that the magnetic field strength  has a power--law
increase, \ie\ $B \propto n({\rm H_2})^{\kappa}$, then we can  use the densities
and field strengths derived in the filament and in the core to derive  the
power--law  index. From the values given in Table~\ref{tab-houde09},  $\kappa\
\simeq 0.2$.  This is significantly lower than the value expected for a  weak
magnetic field with magnetic flux conservation, $\kappa = 2/3$, 
\citep{Crutcher99}. But it is also lower than the value predicted for the
standard  ambipolar diffusion models,  $\kappa \simeq 0.44$--0.5,
\citep{Fiedler93}.  
Ambipolar diffusion appears to be efficient only when densities reach values of 
$\ga 10^8$~\cmt \citep{Tassis07}.

One possibility is that the diffusion arises from fast magnetic reconnection in
the  presence of turbulence \citep{Lazarian99, Santos10}.  Recently,
\citet{Leao12} have carried out simulation to test this scenario in the case of
gravitationally collapsing dense cores with initial turbulent to 
magnetic field energy ratios of 1.6--3. These values are clearly larger than
the  value found in the parsec scale filament, but are only slightly higher than
the value found in the \dr\ core. \citet{Leao12} compute the temporal evolution
of the  average magnetic field--to--density ratio at the radius 0.3 pc in the
core  ($B_{0.3}/\rho_{0.3}$) normalized  by the average value over the entire
cloud, $(B/\rho$), which is 3.2~pc in their computation.  We compute this value
for  the \dr\ core, normalized by the average value in the  filament. From 
Table~\ref{tab-houde09}, we obtain 
$(B_{\rm core}/\rho_{\rm core})/(B_{\rm filament}/\rho_{\rm filament}) \simeq 0.042$.  
Note that the scales used here are 
different with respect to the ones used by \citet{Leao12}.  The \dr\ core has
a radius of 0.04~pc, whereas the filament has an average radius  of $\sqrt{4
\times 0.34}=1.2$~pc (0.34 and 4~pc are roughly the length and  thickness of the
filament). Thus the scales and the scale ratio used in the \dr\ region are
different from the values used in the simulations.  Nevertheless,  the derived
value can be used as a qualitative comparison between the simulations and our
observations. \citet{Leao12} find that in most cases, the aforementioned
parameter is in the 0.1--0.3 range. Lower values are obtained only when Ohmic
dissipation is included.  Therefore, it is a feasible mechanism in the core.

\subsection{Angular Momentum versus Magnetic Fields}

The emission of the \htcop\ 4--3 line associated with the \dr\ dense core shows
a clear velocity pattern along the NE--SW direction (see Figures \ref{Fh13co+} and
\ref{Fpv}). This velocity gradient agrees with the one observed in the \htcop\ 1--0
and N$_2$H$^+$ 1--0 line emission around the core \citep{Schneider10}. These
two lines trace well the large scale filament kinematics.  These lines show an 
interesting  E--W gradient with direction reversals along the filament. 
\citet{Schneider10} interpret the velocity pattern in the filament as evidence of
converging flows, which would have formed the filament.  However, as already 
introduced in Section~\ref{Skin}, the global kinematics in the core can be explained as 
Keplerian--like rotation. We speculate that the observed rotation has been
induced by the large scale motions. The Keplerian--like rotation breaks 
in the inner  region of the core, where the 
hottest and most massive condensations,
SMA~6 and SMA~7, are located. Figure~\ref{Fpv} shows that the Keplerian velocity
distribution in the position--velocity cut along the major axis for a dynamical
mass of  $10 \, {\rm M_{\odot}}/\cos(i)$ ($i$ is the inclination angle of the
rotation axis in the plane of the sky) matches well the velocity pattern of the
\htcop\ 4--3 emission.  This mass is a lower limit, and, indeed, we can estimate
how much mass is  embedded in the inner part of \dr\ assuming that the center of
the core is dominated  by SMA~6 and SMA~7. \citet{Zapata12} estimates that the
mass of these two hot  cores is $\simeq 47$~\msun. If we consider  the total
luminosity of $\simeq 1 \times 10^4$~\lsun in  MM~1 to originate mainly from the
two hot cores,   then it is reasonable to assume that these two  sources harbor
protostars with  a mass of at least $\simeq 13$~\msun.   This yields a total
mass of $\ga 60$~\msun.  This suggests a rotation axis of the core that is
almost along the line-of-sight with  $i \ga 80\arcdeg$.  Since a rotating
envelope is expected to be somewhat flattened in the plane perpendicular to the
rotation's axis, this result suggests that \dr\ is nearly face-on. This could explain 
the lack of flatness observed in the emission of both the dust
continuum and of the \htcop\ 4--3 emission. 

\citet{Machida05} show that the importance of the angular momentum  with respect
to the magnetic fields can be measured from the ratio between the angular
velocity and the magnetic field strength, $\omega/B$, with a critical value 
given by $(\omega/B)_{\rm crit} = 3.19 \times 10^{-8} \,  c_{\rm
s}^{-1}$~yr$^{-1}$~$\mu$G$^{-1}$, where $c_{\rm s}$ is the sound speed in \kms.
The sound speed for the  temperature in the core, 30~K, is  $c_{\rm
s}=0.33$~\kms, so the critical value is $9.8 \times 10^{-8}$~yr$^{-1}$~$\mu$G$^{-1}$. 
We can derive the angular velocity from Figure~\ref{Fpv}, adopting the core's radius, 
$5\farcs2$ (7800~AU). At this radius, the rotation velocity component along the line of 
sight is 1.1~\kms.
Taking into account that the rotation axis has an inclination  of $80\arcdeg$,
the rotation velocity  is $\simeq6.3$~\kms. For the adopted radius, this yields
an angular velocity of $\simeq 2\times10^{-4}$~yr$^{-1}$ and an angular
velocity-to-magnetic strength ratio of $\omega/B \simeq 8.4\times 10^{-8}$, 
which is similar to the critical value.  This suggests that in \dr\ the 
centrifugal  energy is dynamically as important as the magnetic energy. 

Going back to the magnetic field, to better understand its morphology, we have 
to take into account that we are looking at \dr\ in a face-on projection. 
Theoretical models of rotating and magnetized envelopes show that, in a face-on
configuration a spiral magnetic field pattern is expected if initially the
rotation and magnetic field axes are aligned \citep[e.g.,][]{Machida05, 
Padovani12, Kataoka12}.  However, if this is not the case, a more complex
morphology would be expected \citep{Machida05, Hennebelle09}.  Therefore, the
complex polarization pattern observed with the SMA in  \dr\ is probably due  to
the face-on orientation of magnetic field lines that are being wrapped and
twisted by the core's rotation. Indeed, we can estimate the angle of the average
field with respect to the plane of the sky, $\alpha = \arctan(B_{\rm los}/B_{\rm
pos})$. Using the values obtained from the CN Zeeman observations
\citep{Crutcher99} and  from the SMA dust polarization (see Section~\ref{ana}), the
mean magnetic field has an inclination of only $\simeq 10\arcdeg$ with respect
to the plane of the sky.  This suggest that it has a toroidal configuration, which
supports the evidence that the field lines are being dragged by the rotation. 
Simulations show that under these circumstances, the magnetic field tension
would create a large-scale tower low-velocity outflow perpendicular to the 
flattened structure \citep{Tomisaka98, Peters11}. It is possible that the
large-scale  NW--SE, low-velocity CO outflow detected with the JCMT
\citep{Vallee06} is tracing this predicted tower flow.  Note that this outflow
is  not detected with the SMA, suggesting that it has a wide origin.

A final issue about the angular momentum is that SMA~6 and SMA~7 appear to
clearly depart from the Keplerian rotation, because from the hot-core lines the
mean velocity is about 2\kms\ lower than the value expected for Keplerian
rotation velocities (see the bottom panel of Figure~\ref{Fpv}). One possible
explanation for this apparent lack of angular momentum conservation would be
magnetic braking, which have already been observed in another massive  dense
core \citep{Girart09}. However, an alternative possibility is that the angular
momentum have been transferred into the formation of the two sub cores,  SMA~6 
and SMA~7.

\subsection{The High Level of Fragmentation in \dr} 

Recent simulations of massive dense molecular star-forming cores show  that
magnetic fields and radiative feedback can effectively suppress  fragmentation
\citep{Tilley07, Peters11, Hennebelle11, Myers12}, but that outflow feedback may
promote fragmentation \citep{Wang10}. Observationally,  \citet{Palau13} recently
compiled a list of star forming regions that are in  a very early phase, having
luminosities between few hundreds and  $\sim 10^5$~\lsun, and having millimeter aperture
synthesis observations  with angular  resolutions of  $\la 1000$~AU.  They found
a broad range  of fragmentation,  but with 30\% showing no fragmentation in millimeter
wavelengths. A comparison with  simulations of turbulent and magnetized cores 
\citep{Commercon11}, suggests that the level of fragmentation can be
related to the level of magnetization.  Our SMA observation of \dr\  can be
included in this list. By doing this, this source appears to be in  the extreme
case of fragmentation, with more than 10 millimeter sources  detected. This is a
case similar to OMC-1S-136 \citep{Palau13}.   The cases of high fragmentation
can be explained if the cloud is only very weakly  magnetized, with
mass--to--flux ratios of $\sim 100$ \citep{Commercon11, Myers12}.   However, 
the dust continuum observations show that the mass--to--flux  ratio is $\sim 6$.
Even accounting for the mass already accreted onto the  protostars in the \dr\
core (possibly a factor less than two), the value is still  much lower.

A high angular momentum of the core and the outflow feedback seem to be a 
plausible explanation for the \dr\ fragmentation. First, following  the
\citet{Chen12} recipe, we estimate the rotational energy-to-gravitational 
potential energy ratio for \dr\ to be $\sim 0.5$ for the corrected projection
of  the rotation velocity.    This value is an order of magnitude higher than
the values reported in the  \citet{Palau13} survey, although the values derived
in this survey were  uncorrected for projection. However, since it is expected
that this sample has a random distribution of source orientations, we can
consider that \dr\ is a core with a significantly higher value of angular
momentum than the average core. Second, this source shows a very active outflow
activity (see Section~\ref{co}),  with emission from high density tracers
\citep{Zapata12} in the outflows and the  presence of a rich variety of masers
(see Section \ref{intro} for references).

\section{Conclusions}

We have carried out an extensive molecular, dust and polarimetric study of the 
massive \dr\ star forming core from SMA high-angular-resolution observations at
880~$\mu$m. We have obtained observations from all the available SMA 
configurations (subcompact, compact, extended and very extended).  We have  also
included complementary archival polarimetric observations from SCUPOL of the
JCMT telescope \citep{Matthews09}.  All these data allow us to study and
characterize the magnetic field properties from parsec scales down to 1000~AU
toward a core that appears to be highly fragmented at smaller scales. The
molecular line emission of selected transition detected with the SMA allows us
to study the kinematic properties of the core and to put them into a context
together  with the magnetic field properties. Here, we summarize the main
results:

\begin{enumerate}

\item The SMA maps at different angular resolutions ($3\farcs6$-$0\farcs75$)
reveal a complex magnetic field morphology in the \dr\ core. This is in
contrast  to the relatively smooth large-scale magnetic field threading the 
filamentary dense ridge where \dr\ is embedded. 

\item The $\sim 7.8$~GHz bandwidth reveals a rich molecular line spectra in
the \dr\ core. In particular, SMA 6 and SMA 7 have  spectral features consistent
with  being hot molecular cores. The \htcop\ 4--3 emission correlates well with 
the dust emission at scales of 0.01--0.1~pc except toward the SMA~6 and 
SMA~7 hot cores.  Therefore, this line is a good tracer of the overall \dr\ 
core's kinematics. 

\item Combining the kinematic information from selected molecular tracers
(\htcop\ 4--3 that traces the \dr\ core except SMA~6 and SMA~7, ethyl cyanide
tracing  SMA~6, dimethyl ether tracing SMA~7), we find that the \dr\ kinematics
are compatible with Keplerian motions, except in the center of the core around 
SMA~6 and SMA~7. From the mass enclosed in SMA~6 and SMA~7, we estimate 
that the rotation axis is close to the line-of-sight. The \dr\ core is, thus, probably
observed face-on. 

\item The HV CO 3--2 emission shows two collimated bipolar outflows
approximately in the east-west direction, $PA \simeq  65\arcdeg$ and 
$110\arcdeg$. They are probably powered by SMA~7 and SMA~4, respectively.  
SMA 6 also powers a compact outflow in the N--S direction \citep{Minh11}.

\item The statistical analysis reveals that the magnetic field is approximately
in equipartition with the turbulent energy in  the \dr\ core, whereas in the
filament the magnetic field energy dominates over turbulence. This possibly 
suggests that the star formation activity (for example through the powerful
outflows)  is injecting  turbulence in the \dr\ core. This analysis in \dr\
yields a turbulent  length scale, 16~mpc, and a magnetic field component in the
plane of  the sky, 2.1 mG. These values are in good agreement with the
values  derived from a completely independent method by \citet{Hezareh10}. 

\item The total magnetic field strength derived, combining the dust
measurements  with previous Zeeman measurements \citep{Crutcher04}, is 2.1 and
0.9~mG for  the \dr\ core and the parsec-scale filament, respectively. Both
molecular structures  are supercritical, in agreement with the observed
large-scale infall motions  \citep{Schneider10}.  The mass--to--flux ratios for
the core and the ridge are 5.9 and  3.4 times the critical value, respectively. 
Thus, gravity has overcome the interstellar magnetohydrodynamic turbulence and the 
magnetic fields threading both the DR~21 filamentary ridge  and especially the \dr\
core.  An independent analysis based on the polarization--intensity gradient
method \citep{koch12a} further confirms this finding with a map-averaged
field-to-gravity force ratio of about 0.8, and some local areas where the field
significance is  reduced to $\sim$10\% or less. The magnetic field direction 
has an inclination of only $\simeq10\arcdeg$ with respect to the plane of the sky,
suggesting a toroidal configuration. 

\item In spite of being clearly supercritical, the high fragmentation observed
in \dr\ would require a much higher mass--to--flux ratio according to recent 
simulations of turbulent and magnetized clouds \citep{Commercon11, Myers12}. It
is possible that the high angular momentum measured is playing an important 
role  in the fragmentation process of the \dr\ core. First, the ratio between 
the angular velocity and the magnetic flux, $\omega/B$, is similar to the 
critical value, indicating that rotation is energetically as important as the
magnetic fields in the dynamics of the core. This can explain the  toroidal
configuration of the magnetic field lines, as they are being wrapped by the 
rotation of the dense gas. 

\item The wrapped and toroidal magnetic field configuration suggests that the 
previously reported large-scale low-velocity  CO outflow \citep{Vallee06}, 
undetected with the SMA, is tracing the theoretically predicted large-scale
tower flow  \citep{Peters11}.

\end{enumerate}

\acknowledgments
The SMA data were taken as part of the Legacy SMA project ''Filaments, Star 
Formation and Magnetic Fields'' (PI: Qizhou Zhang).  The Submillimeter Array 
is a joint project between the Smithsonian Astrophysical Observatory and the 
Academia Sinica Institute of Astronomy and Astrophysics and is funded by the 
Smithsonian Institution and the Academia Sinica. We thank all members of the
SMA staff that made these observations possible,  as well as the ASIAA and the
SAO for the support on the Legacy project.  We also thank M. Padovani for 
providing the polarization data from simulations. JMG and PF are supported by the
Spanish MINECO AYA2011-30228-C03-02, AYA2008-04451-E and Catalan AGAUR 
2009SGR1172 grants. S.P.L. acknowledges support from the National Science 
Council of Taiwan with grants NSC 98-2112-M-007- 007-MY3 and 
NSC 101-2119-M-007-004.

\end{document}